\documentclass[jcp,aip,reprint]{revtex4-1}
\usepackage{amsmath}
\usepackage{graphicx}
\usepackage{epstopdf}
\makeatletter
\g@addto@macro\bfseries{\boldmath}
\makeatother

\begin{document}

\title{Observation of b$_2$ symmetry vibrational levels of the SO$_2$ $\tilde{\mbox{C}}$ $^1$B$_2$ state: Vibrational level staggering, Coriolis interactions, and rotation-vibration constants}
\author{G.\ Barratt Park}
\email{barratt.park@mpibpc.mpg.de}
\affiliation{Department of Chemistry, Massachusetts Institute of Technology, Cambridge, Massachusetts 02139}
\affiliation{Current address: Institute for Physical Chemistry, University of G\"{o}ttingen, Germany}
\author{Jun Jiang}
\affiliation{Department of Chemistry, Massachusetts Institute of Technology, Cambridge, Massachusetts 02139}
\author{Catherine A.\ Saladrigas}
\author{Robert W.\ Field}
\email{rwfield@mit.edu}
\affiliation{Department of Chemistry, Massachusetts Institute of Technology, Cambridge, Massachusetts 02139}

\begin{abstract}
The $\mathrm{\tilde{C}}$ $^1$B$_2$ state of SO$_2$ has a double-minimum potential in the antisymmetric stretch coordinate, such that the minimum energy geometry has nonequivalent SO bond lengths. However, low-lying levels with odd quanta of antisymmetric stretch (b$_2$ vibrational symmetry) have not previously been observed because transitions into these levels from the zero-point level of the $\mathrm{\tilde{X}}$ state are vibronically forbidden. We use IR-UV double resonance to observe the b$_2$ vibrational levels of the $\mathrm{\tilde{C}}$ state below 1600 cm$^{-1}$ of vibrational excitation. This enables a direct characterization of the vibrational level staggering that results from the double-minimum potential. In addition, it allows us to deperturb the strong $c$-axis Coriolis interactions between levels of a$_1$ and b$_2$ vibrational symmetry, and to determine accurately the vibrational dependence of the rotational constants in the distorted $\mathrm{\tilde{C}}$ electronic  state. 
\end{abstract}

\maketitle

\section{Introduction}
The $\tilde{\mbox{C}}$ $^1$B$_2$ state of SO$_2$, with origin at $v_{00}=42,573.45$ cm$^{-1}$, has been the subject of extensive research. Early investigations by Duchesne and Rosen\cite{duchesnerosenSO2} assigned the $\tilde{\mbox{C}}\leftarrow \tilde{\mbox{X}}$ band system to two overlapping electronic transitions, but subsequent work by Jones and Coon\cite{CoonSO2} and by Brand and coworkers\cite{Brand_SO2_Cstate,BrandSO2MolPhys} provided evidence that the band system was due to a single electronic transition with unusual vibrational structure.  Most notably, the antisymmetric stretching mode $\nu_3$ of the $\tilde{\mbox{C}}$ state appeared to have an uncharacteristically low fundamental vibrational frequency ($\sim$212 cm$^{-1}$), and to display an unusually large degree of anharmonicity. Further work by Brand and coworkers\cite{Brand_SO2_Cstate,BrandSO2MolPhys} and Hallin and Merer\cite{HallinThesis} provided evidence that the $\tilde{\mbox{C}}$ state has a double minimum potential in $q_3$, with two equivalent minimum-energy configurations that exhibit unequal bond lengths. 

Considerable evidence (including band origin isotope shifts, inertial defects, and centrifugal distortion constants) suggests staggering in the $\nu_3'$ progression and an anomalously low $\nu_3$ fundamental frequency.\cite{Brand_SO2_Cstate} However, transitions to vibrational levels of the $\mathrm{\tilde{C}}$ state with odd quanta of $v_3$ (b$_2$ vibrational character) are forbidden from the ground vibrational level of the $\tilde{\mbox{X}}$ state. Thus, direct observation of the staggering of low-lying levels with even vs.\ odd quanta of $v_3'$ has not previously been reported. The locations of some of the dark b$_2$ vibrational levels had been inferred by fitting the strong $c$-axis Coriolis perturbations between the b$_2$ levels and the bright a$_1$ vibrational levels.\cite{Brand_SO2_Cstate,BrandSO2MolPhys,HallinThesis,SO2_Yamanouchi} Ivanco made an assignment of the $\nu_3'$ fundamental in low-resolution hot laser-induced fluorescence spectra,\cite{IvancoThesis} but our work shows his assignment to be incorrect.

Recently, we analyzed the $c$-axis Coriolis perturbation in a vibrationally-excited level of the $\tilde{\mbox{C}}$ state at 45,335 cm$^{-1}$---approximately 500 cm$^{-1}$ below the dissociation limit---and we made rotationally-resolved observations of the nominally dark perturbing level that borrows intensity via the Coriolis interaction.\cite{MOMplex} To our knowledge, this was the first direct high-resolution measurement on a b$_2$ vibrational level in the $\tilde{\mbox{C}}$ state. However, this level lies $\sim$2756 cm$^{-1}$ above the  $\tilde{\mbox{C}}$-state origin, too high in energy to provide meaningful information about the asymmetry in the potential energy surface that gives rise to a$_1$/b$_2$ level staggering near the bottom of the well. In the current work, we report high resolution spectra of b$_2$ vibrational levels of the $\tilde{\mbox{C}}$ state below $\sim$1600 cm$^{-1}$ of vibrational excitation, which we have recorded using IR-UV double-resonance fluorescence spectroscopy.

This is the first of a three-part series. In this part, we report the staggered vibrational energy levels of the SO$_2$ $\tilde{\mbox{C}}$ state, we perform a rotational analysis of the strong $c$-axis Coriolis interactions, and we determine the rotational constants and rotation-vibration constants. In part II of the series,\cite{SO2_IRUV_2} we fit the available data on the $\tilde{\mbox{C}}$ state (including rotational constants, rovibrational information, and isotopologue data) to an internal coordinate force field model. Using a two-step diagonalization procedure, we characterize a large number of highly mixed levels below 3000 cm$^{-1}$ of vibrational excitation, and we calculate the anharmonic Franck-Condon factors for the $\tilde{\mbox{C}}\leftarrow\tilde{\mbox{X}}$ absorption spectrum. We characterize anharmonically-induced interference effects in the vibronic transition intensities, similar to those recently characterized in the $\tilde{\mbox{A}}\rightarrow\tilde{\mbox{X}}$ emission spectrum of acetylene.\cite{CartesianModel} In part III of the series,\cite{SO2_IRUV_3} we discuss the vibronic mechanisms for the distortion in the SO$_2$ $\tilde{\mbox{C}}$ state. We propose a three-state interaction model, which suggests that the observed level pattern is sensitive to interaction with both a higher bound electronic state and a repulsive state. Furthermore, we model the dependence of level staggering on the bystander mode $\nu_2'$, which is sensitive to the location of a conical intersection between the $\tilde{\mbox{C}}$ state and the bound 2\,$^1$A$_1$ state.

Throughout the current work, we will use the notation $(v_1, v_2, v_3)$ for the vibrational quantum numbers in the normal mode basis. In part II of this series, quantum numbers are given in a basis developed by Kellman and co-workers\cite{KellmanFermiAssign} for Fermi-interacting systems. Because of the strong 1:33 Fermi interaction, this basis provides a more accurate representation of the $\mathrm{\tilde{C}}$-state vibrational levels. However, for the low-lying vibrational states discussed here, there is a one-to-one correspondence between the normal mode labels and the Kellman-type mode labels, so we use normal mode labels for simplicity. 

\section{Selection rules}
The $\tilde{\mbox{C}}$ state of SO$_2$ exhibits a small barrier at the $\mathrm{C_{2v}}$ geometry along the $q_3'$ antisymmetric stretch coordinate. Thus there are two equivalent minimum-energy configurations with $\mathrm{C_s}$ geometry. Despite this fact, it is convenient to treat the molecule in $\mathrm{C_{2v}}$ symmetry, in keeping with previous investigators. The barrier at $\mathrm{C_{2v}}$ is small and is an even function of $q_3$, so there are no terms in the vibrational Hamiltonian that break the vibrational symmetries in the representations of $\mathrm{C_{2v}}$. Therefore, the low-lying levels of the $\mathrm{\tilde{C}}$ state conserve $\mathrm{a_1}$ (even quanta of $v_3$) and $\mathrm{b_2}$ (odd quanta of $v_3$) vibrational symmetries, and the selection rule in the $\mathrm{\tilde{C}}\leftrightarrow\mathrm{\tilde{X}}$ spectrum is such that the vibrational symmetry remains unchanged.

Two mechanisms could plausibly destroy the vibrational selection rules. First, it is possible for vibronic interactions to cause vibrational dependence of the electric dipole transition moment (Herzberg-Teller coupling). Alternatively, $c$-axis Coriolis interaction in the $\mathrm{\tilde{C}}$-state can cause mixing between rovibrational levels that differ in vibrational symmetry. Both effects have been invoked in the interpretation of dispersed fluorescence experiments from highly excited predissociated vibrational levels of the $\mathrm{\tilde{C}}$ state between 210--205 nm, from which the dispersed fluorescence includes transitions to both a$_1$ and b$_2$ vibrational levels of the ground electronic state.\cite{Yang1991,SO2ResonanceEmission, ButlerSO2Emission, Parsons2000499, GuoSO2EmissionSpectra} Brand \textit{et al.}\cite{SO2ResonanceEmission} propose that the most likely mechanism is Coriolis interaction, but Ray \textit{et al.}\cite{ButlerSO2Emission} argue that the cold rotational temperature of their supersonic expansion rules out Coriolis interactions. In the current work, we observe nominally forbidden transitions that borrow intensity via Coriolis interactions at rotational quantum numbers as low as $J=2$, so we do not believe that it is possible to rule out Coriolis interactions without first understanding the level structure. However, in the 210--205 nm region, theory predicts an avoided crossing between the $\mathrm{\tilde{C}}$ state and the repulsive 3 $^1$A$'$ state.\cite{19912792,Katagiri_SO2_photodissoc,Bludsk2000607} Therefore, it is possible that both rotation-vibration and vibration-electronic interactions give rise to $\mathrm{a}_1/\mathrm{b}_2$ admixture in the 210--205 nm region. However, in the low-lying non-predissociated region below 2000 cm$^{-1}$ of vibrational excitation, there is no evidence for vibronically allowed transition intensity. In the absence of rovibrational Coriolis interactions, the observed transitions strictly conserve the vibrational symmetry.

The $\mathrm{\tilde{C}}$ $\mathrm{^1B_2}$ $\leftarrow\mathrm{\tilde{X}}$ $\mathrm{^1A_1}$ transition has an $a$-axis electronic transition moment, and obeys $a$-type rotational selection rules. However, as a consequence of oxygen atom nuclear spin statistics, only half of the rovibronic levels exist. The total wavefunction must be even with respect to interchange of the spin-zero oxygen nuclei. The nuclear spin component of the wavefunction is, by necessity, even. Table \ref{C2v} summarizes the representations of the rotational and vibrational wavefunctions of SO$_2$ in the C$_{\mathrm{2v}}$ molecular group. Since the equivalent rotation for the (12) operation that exchanges the oxygen nuclei is $C_{2b}$, the total rotation-vibration-electronic wavefunction must therefore be either $\mathrm{A_1}$ or $\mathrm{A_2}$. In the I$^{\mathrm{r}}$ representation of the rotational wavefunctions, even (e) and odd (o) values of $K_aK_c$ correspond to the $\mathrm{C_{2v}}$ molecular group representations as follows: $\Gamma\mathrm{_{rot}(ee)= A_1}$, $\Gamma\mathrm{_{rot}(oo)= A_2}$, $\Gamma\mathrm{_{rot}(eo)= B_1}$, and $\Gamma\mathrm{_{rot}(oe)= B_2}$. As a result, levels with A$_1$ or A$_2$ vibration-electronic character may only have ee or oo rotational wavefunctions and levels with B$_1$ or B$_2$ vibration-electronic character may only have eo or oe rotational wavefunctions. 
The consequence is that in the absorption spectrum from the a$_1$ ground vibrational state, the combination of vibrational selection rule and $a$-type rotational selection rule leads to $\mathrm{a_1(eo)'\leftarrow a_1(ee)''}$ or $\mathrm{a_1(oe)'\leftarrow a_1(oo)''}$ transitions, whereas in IR-UV experiments using a b$_2$ intermediate vibrational level, the transitions are of the type $\mathrm{b_2(ee)'\leftarrow b_2(eo)''}$ or $\mathrm{b_2(oo)'\leftarrow b_2(oe)''}$. (Here, the a$_1$ and b$_2$ labels refer only to the symmetry of the vibrational part of the wavefunction.) When rovibrational levels of the $\tilde{\mbox{C}}$ state are mixed via $c$-axis Coriolis interactions, intensity borrowing gives rise to nominally forbidden transitions of the type $\mathrm{b_2(oo)'\leftarrow a_1(ee)''}$, $\mathrm{b_2(ee)'\leftarrow a_1(oo)''}$, $\mathrm{a_1(oe)'\leftarrow b_2(eo)''}$ or $\mathrm{a_1(eo)'\leftarrow b_2(oe)''}$.
\begin{table}
\caption{The character table for the C$_{\mathrm{2v}}$ molecular group. The symmetries of the vibrational and rotational parts of the wavefunction are given on the right side of the table. The molecular frame axes are specified using principal inertial axis labels.  \label{C2v}} 
\centering
\begin{tabular}{r | cccc c | cc} \toprule
$\mathbf{C_{2v}(M)}$: & $E$ & (12)           & $E^*$                  & (12)$^*$ &&&\\
$\mathbf{C_{2v}}$:       & $E$ & $C_{2b}$ & $\sigma_{ab}$ & $\sigma_{bc}$ &&&\\
Equiv.\ rot:                       & $E$ & $C_{2b}$ & $C_{2c}$ & $C_{2a}$ &              &$\psi_{\mathrm{vib}}$& $K_aK_c$ 
\\ \colrule
A$_1$                             & 1      & $\phantom{-}1$&$\phantom{-}1$&$\phantom{-}1$ :& $T_b$       &$\nu_1,\nu_2$             & $ee$ \\
A$_2$                             & 1      & $\phantom{-}1$&$-1$                    & $-1$ :                   & $J_b$        &                                       & $oo$ \\
B$_1$                             & 1      & $-1$                   &$-1$                  & $\phantom{-}1$ :  & $T_c,J_a$&                                       & $eo$ \\
B$_2$                             & 1      & $-1$                   &$\phantom{-}1$ & $-1$ :                   & $T_a,J_c$&$\nu_3$                        & $oe$ \\
\botrule                         
\end{tabular}
\end{table}

\section{Experimental}\label{experiment}
\subsection{Strategies for observing the b$_2$ levels}\label{Strategies}
Prior to our IR-UV double resonance experiments, we attempted other schemes to observe the $\nu_3'$ fundamental of the $\tilde{\mbox{C}}$ state, including hot-band pumping, using a vibrationally hot (but rotationally cold) expansion from a heated nozzle. This approach proved to be difficult, because, although the antisymmetric stretching mode $\nu_3$ has the \emph{lowest} fundamental vibrational frequency in the $\tilde{\mbox{C}}$ state (212 cm$^{-1}$), it has the \emph{highest} fundamental vibrational frequency in the $\tilde{\mbox{X}}$ state (1362 cm$^{-1}$). The maximum population in the $\tilde{\mbox{X}}(0,0,1)$ level does not occur until a vibrational temperature of $\sim$1200 K is reached, at which temperature the population in $\tilde{\mbox{X}}(0,0,1)$ is only 5\%. It is even more unfortunate that the much stronger $\tilde{\mbox{C}}(0,0,0)\leftarrow\tilde{\mbox{X}}(1,0,0)$ hot band lies within 2 cm$^{-1}$ of the desired $\tilde{\mbox{C}}(0,0,1)\leftarrow\tilde{\mbox{X}}(0,0,1)$. Thus, it was difficult to identify hot band transitions to the $\tilde{\mbox{C}}(0,0,1)$ level. IR-UV double resonance proved to be a much more effective technique. We chose the $\tilde{\mbox{X}}(1,0,1)$ state as the IR intermediate because of the strength of the $\tilde{\mbox{X}}(1,0,1)\leftarrow\tilde{\mbox{X}}(0,0,0)$ IR transition at 2500 cm$^{-1}$.

\subsection{Experimental details}
The experimental apparatus for the IR-UV measurements has been described previously,\cite{FASE} so we give only a brief summary of our apparatus, but we expand on details unique to the SO$_2$ experiment. Tunable IR radiation was produced by difference frequency generation in a LiNbO$_3$ crystal pumped by an injection-seeded Nd:YAG laser (Spectra-Physics PRO-270) at 1064 nm and a tunable dye laser (Lambda Physik FL2002) operating at 840 nm (LDS 821 dye), which was pumped by the 2nd harmonic (532 nm) output of the same Nd:YAG laser. We obtained an IR pulse energy of $\sim$400 $\mu$J at 2500 cm$^{-1}$. To ensure resonance of the IR frequency with transitions in the $\tilde{\mbox{X}}(1,0,1)\leftarrow\tilde{\mbox{X}}(0,0,0)$ band, absorption signals were monitored in a photoacoustic cell containing 15 Torr of neat SO$_2$ at room temperature. 

The grating-limited IR spectral width of 0.1 cm$^{-1}$ is not sufficiently narrow to ensure selection of only a single rotational eigenstate as the IR-UV intermediate. Thus, our double resonance spectra contain features from intermediate levels populated via several nearby IR transitions. We used the relatively sparse R branch of the $\tilde{\mbox{X}}(1,0,1)\leftarrow\tilde{\mbox{X}}(0,0,0)$ transition to minimize excitation of these ``extra'' intermediate levels, but typically more than one $K_a$ rotational level was excited by the IR laser. We used this to our advantage to collect double resonance spectra via the various $K_a$ sub-bands simultaneously. For each $J$, the IR laser was typically broad enough to excite transitions with $K_a=$0--3.  Because the IR spectrum is well understood, it was straightforward to predict the $K_a$ levels as well as any other spurious levels that were populated by nearby IR transitions. An overview of transitions that were excited in our typical IR pump schemes may be found in Table S.II of the Supplementary Material.\cite{SO2_IRUV_1_Supplement}

To generate the UV photon, the 355 nm third harmonic of the same Nd:YAG laser used in the IR generation pumped a second dye laser (Lambda Physik FL3002E) to produce tunable laser radiation over the range 480--497 nm (Coumarin 480 or 503). This output was frequency doubled in a $\beta$-barium borate crystal and a small portion of the fundamental was passed through a heated $^{130}$Te$_2$ vapor absorption cell for frequency calibration. An intracavity etalon reduced the spectral width to 0.04 cm$^{-1}$, and, after frequency doubling, the UV power was approximately 100--200 $\mu$J/pulse. The IR and UV beams were counter-propagated through a molecular beam chamber. As a single Nd:YAG laser generated both pulses, their relative arrival times at the chamber could be controlled only by adding a delay line to the UV beam path. The length of the delay line was chosen such that the UV pulse arrived 15 ns after the IR pulse. 

The IR and UV pulses interacted with an unskimmed supersonic jet of 0.1\% SO$_2$ in He, expanded through a General Valve (Series 9, $d=1.0$ mm). The jet was backed by a pressure of 1 atm, and the chamber operated at $\sim$1$\times10^{-5}$ Torr average pressure while under gas load. The IR and UV radiation intersected the jet at a distance of 2 cm from the nozzle. A Hamamatsu R375 photomultiplier tube collected the laser-induced fluorescence at an angle mutually perpendicular to the laser path and molecular beam, using $f$/1.2 collection optics and a UG-11 filter to block laser scatter. The IR and UV lasers were sent through a set of baffles, described in Ref.\ \onlinecite{FASE}, to minimize light scattered onto the detector. The photomultiplier tube signal was split and one line was input to a 30 dB low-noise voltage amplifier (Femto DHPVA-200) to increase the sensitivity and dynamic range of detection. The fluorescence decay was recorded on an oscilloscope and had a lifetime of typically 40 ns. For each $\sim$0.018 cm$^{-1}$ frequency resolution element, the fluorescence signal was averaged for 20 laser shots. The fluorescence spectrum was obtained by integrating the first 30 ns of fluorescence decay. 

Table S.I of the supplementary material summarizes the term values for rovibronic levels of the SO$_2$ $\mathrm{\tilde{C}}$ state observed in the current work.\cite{SO2_IRUV_1_Supplement}

\section{Rotational structure and Coriolis interactions}
Because the effective $\nu_3$ frequency is depressed in the $\mathrm{\tilde{C}}$ state (particularly at low vibrational excitation), it is brought into near resonance with $\nu_2$. This results in closely spaced sets of levels with conserved quanta of $v_2+v_3$, which are coupled via $c$-axis Coriolis interactions, resulting in admixture of $\mathrm{a}_1$ and $\mathrm{b}_2$ vibrational characters. In the basis of harmonic vibrational wavefunctions and signed-$k$ symmetric top rotational wavefunctions, the form of the resulting lowest-order matrix element is 
\begin{multline}\label{Coriolis23_1}
\langle v_1,v_2-1,v_3+1,J,k\pm1 | \mathbf{H} |v_1,v_2,v_3,J,k\rangle  = \\ 
\mp C\zeta_{23}^{(c)}\Omega_{23}[J(J+1)-k(k\pm1)]^{1/2}[v_2(v_3+1)]^{1/2},
\end{multline}
where $\Omega_{kl}=1/2[(\omega_k/\omega_l)^{1/2}+(\omega_l/\omega_k)^{1/2}]$ and $\zeta_{23}^{(c)}\approx0.93$ in the harmonic approximation. $c$-axis Coriolis interactions between $\nu_1$ and $\nu_3$ are also possible:
\begin{multline}\label{Coriolis13}
\langle v_1-1,v_2,v_3+1,J,k\pm1 | \mathbf{H} |v_1,v_2,v_3,J,k\rangle= \\
\mp C\zeta_{13}^{(c)}\Omega_{13}[J(J+1)-k(k\pm1)]^{1/2}[v_1(v_3+1)]^{1/2},
\end{multline}
and the sum rule $|\zeta_{23}^{(c)}|^2+|\zeta_{13}^{(c)}|^2=1$ applies. However, this latter interaction is relatively unimportant because the energy denominator for the interaction is large. The matrix elements (\ref{Coriolis23_1})--(\ref{Coriolis13}) are added to the rigid rotor Hamiltonian matrix elements in the same basis,
\begin{multline}
\langle v_1,v_2,v_3,J,k|\mathbf{H}| v_1,v_2,v_3,J,k\rangle \\
=T_0(v_1,v_2,v_3)+\left[A-\frac12(B+C)\right]k^2  \\
+\frac{1}{2}(B+C)J(J+1),
\end{multline}
\begin{multline}
\langle v_1,v_2,v_3,J,k\pm2|\mathbf{H}|v_1,v_2,v_3,J,k\rangle \\
=\frac{1}{4}(B-C)[J(J+1)-k(k\pm1)]^{1/2} \\
\times[J(J+1)-(k\pm1)(k\pm2)]^{1/2},
\end{multline}
and the resulting matrix is diagonalized to obtain the rovibrational energies. Because the $\mathrm{\tilde{C}}$ state is highly anharmonic, we do not attempt to describe the interactions in terms of a global $\zeta_{23}^{(c)}$ parameter. Instead, we assign each set of levels an effective interaction strength $t_1^{(n)}$, which replaces $C\zeta_{23}^{(c)}\Omega_{23}[v_2(v_3+1)]^{1/2}$ in Eq.\ (\ref{Coriolis23_1}) to give matrix elements of the form
\begin{multline}\label{Coriolis23}
\langle v_1',v_2',v_3',J,k\pm1 | \mathbf{H} |v_1,v_2,v_3,J,k\rangle \\= 
\mp t_1^{(n)} [J(J+1)-k(k\pm1)]^{1/2}.
\end{multline}
Following the notation of Ref.\ \onlinecite{HallinThesis}, we use $t_m^{(n)}$ for the rotationally independent prefactor of the vibration-rotation matrix element, where $m$ gives the power of the rotational operator(s), $J_{\alpha}$, and $n$ gives the combined power of the vibrational momentum and position operators, $p_k$ and $q_{k}$, in the matrix element. In fitting our IR-UV data, acquired under supersonic jet expansion conditions with $T_{\mathrm{rot}}\approx10$ K, it was not necessary to include the effects of centrifugal distortion. However, in fits incorporating data from Ref.\ \onlinecite{HallinThesis}, acquired at dry ice temperature, we include quartic centrifugal distortion terms, using Watson's A reduction in the I$^{\mathrm{r}}$ representation.\cite{WatsonHamiltonian}

Table \ref{effrotconstants} lists the effective rotational constants obtained from a fit to each vibrational level individually, ignoring Coriolis interactions. As noted by Hallin\cite{HallinThesis} and Yamanouchi \textit{et al.},\cite{SO2_Yamanouchi} many of the levels are rotationally perturbed, as suggested by the wide range of $C$ rotational constants, large magnitudes of the inertial defect, and average fit errors much larger than the frequency calibration uncertainty ($\sim$0.02 cm$^{-1}$). A large positive or negative inertial defect, respectively, is a signature of a $c$-axis Coriolis perturbation from a higher lying or lower lying level. Throughout this work, we calculate the lower-state term energies from the parameters of Refs.\ \onlinecite{Helminger198566} and \onlinecite{Pine1977386} for the $\mathrm{\tilde{X}}(0,0,0)$ and $\mathrm{\tilde{X}}(1,0,1)$ levels, respectively.

\begin{table*}
\caption{Effective rotational constants and band origins ($T$) for the vibrational levels of the $\mathrm{\tilde{C}}$ state of SO$_2$ below $\sim$1600 cm$^{-1}$ of vibrational excitation, obtained by fitting the rotational structure of each band separately. All energies are in cm$^{-1}$ units and the inertial defects ($\Delta$) are listed in amu$\cdot$\AA$^2$ units. Values in parentheses give the 2$\sigma$ uncertainty in the final significant digit. \label{effrotconstants}} 
\centering
\begin{tabular}{llllllll} \toprule
$(v_1',v_2',v_3')$&\multicolumn{1}{c}{$T$}&  \multicolumn{1}{c}{$T_{\mathrm{vib}}$}& \multicolumn{1}{c}{$A_{\mathrm{eff}}$} & \multicolumn{1}{c}{$B_{\mathrm{eff}}$} &\multicolumn{1}{c}{$C_{\mathrm{eff}}$}    &\multicolumn{1}{c}{$\Delta$}& \multicolumn{1}{c}{Ave error} \\
\colrule 
(0,0,0)\footnote{Ref.\ \onlinecite{HallinThesis}}  
              & 42573.450(4)  & \multicolumn{1}{c}{0}      & 1.15050(9) & 0.34751(5)   & 0.26537(4)   &$ \phantom{-1}0.361   $&0.008 \\
(0,0,1)  & 42786.026(6)  & \phantom{1}212.576(6)  & 1.1474(16) & 0.3444(5)   & 0.2614(4)     &$  \phantom{-1}0.8620  $& 0.017  \\
(0,1,0)\footnotemark[1]  
              & 42950.933(4)  & \phantom{1}377.483(4)  & 1.17052(6) & 0.34589(4)  & 0.26576(4)   &$ \phantom{-1}0.292   $&0.010  \\
(0,0,2)\footnotemark[1]    
              & 43134.672(7)  & \phantom{1}561.222(4)  & 1.14509(19)& 0.34227(11) & 0.24572(11) &$ \phantom{-1}4.632 $&0.041 \\
(0,1,1)  & 43155.635(8)  & \phantom{1}582.184(8)  & 1.1722(8)     & 0.3443(5)   & 0.2743(7) &$\phantom{1}$$-1.876$&0.016 \\
(0,2,0)\footnotemark[1]    
              & 43324.992(4)  & \phantom{1}751.542(4)  & 1.19140(12)& 0.34429(4)&0.26565(3) &$  \phantom{-1}0.344 $&0.013 \\
(0,0,3)  & 43464.382(6)  &\phantom{1}890.932(6)   & 1.1424(13)   & 0.3407(5)  &0.2498(4)   &$\phantom{-1}3.242    $&0.015   \\
(0,1,2)\footnotemark[1]    
              & 43505.31(10)  &\phantom{1}931.87(1)     & 1.152(14)     & 0.336(4)     &0.242(4)     &$\phantom{-1}4.85       $&0.11   \\
(0,2,1)  & 43522.564(10)&\phantom{1}949.114(10)& 1.1908(20)    & 0.3430(7)   &0.2906(6)   &$\phantom{1}$$-5.29$ &0.036\\
(1,0,0)\footnotemark[1]   
             & 43533.51(12)  &\phantom{1}960.06(12)   &1.149(9)       &0.346(15)     &0.266(16)   & $\phantom{1}$$-0.107$&0.17\\
(0,3,0)\footnote{Ref.\ \onlinecite{SO2_Yamanouchi}}  
             & 43695.48(3)    &1122.03(3)   &1.209(12)    &0.3419(24)   &0.2650(21)  &$\phantom{-1}0.364 $& \\
(0,0,4)\footnotemark[2] 
             & 43818.90(5)    &1245.45(9)    &1.113(	\!\!17)  &0.345(2)        &0.2008(18)   &$\phantom{-}20.01 $& 0.031 \\
(1,0,1) & 43834.762(7)  &1261.311(7) &1.1459(12) &0.3420(3)      &0.2558(3)    &$\phantom{-1}1.913 $& 0.0165  \\
(0,1,3) & 43825.81(5)    &1252.36(5)   &1.209(19)    &0.349(3)        &0.2926(20)  &$\phantom{1}$$-4.631$&0.094\\
(0,2,2)\footnotemark[2] 
            & 43873.42(2)    &1299.97(2)   &1.188(13)    &0.3453(18)   &0.2069(23)  &$\phantom{-}18.47     $&0.017\\
(0,3,1) & 43886.705(21)&1313.255(21)&1.222(4)   &0.3452(17)    &0.3188(14)  &$\phantom{1}$$-9.77$ &0.034\\
(0,0,5) & 44169.266(17)&1595.816(21)&1.129(4)   &0.3411(13)   &0.2128(10)   &$\phantom{-}14.85    $&0.052\\
(0,1,4)\footnotemark[2] 
            & 44177.70(2)    &1604.25(2)    &1.1596(147) &0.3535(16)    &0.1795(11)   &$\phantom{-}31.70   $ &       \\
(1,0,2)\footnotemark[2] 
            & 44227.12(1)    &1653.67(1)   &1.1410(66)&0.3408(13)    &0.2709(7)     &$\phantom{1}$$-2.37$ & \\
\botrule                         
\end{tabular}
\end{table*}

\subsection{The $\mathrm{\tilde{C}}$(0,0,1) level}\label{section001}

\begin{figure*}
\centering
\includegraphics[width=\linewidth]{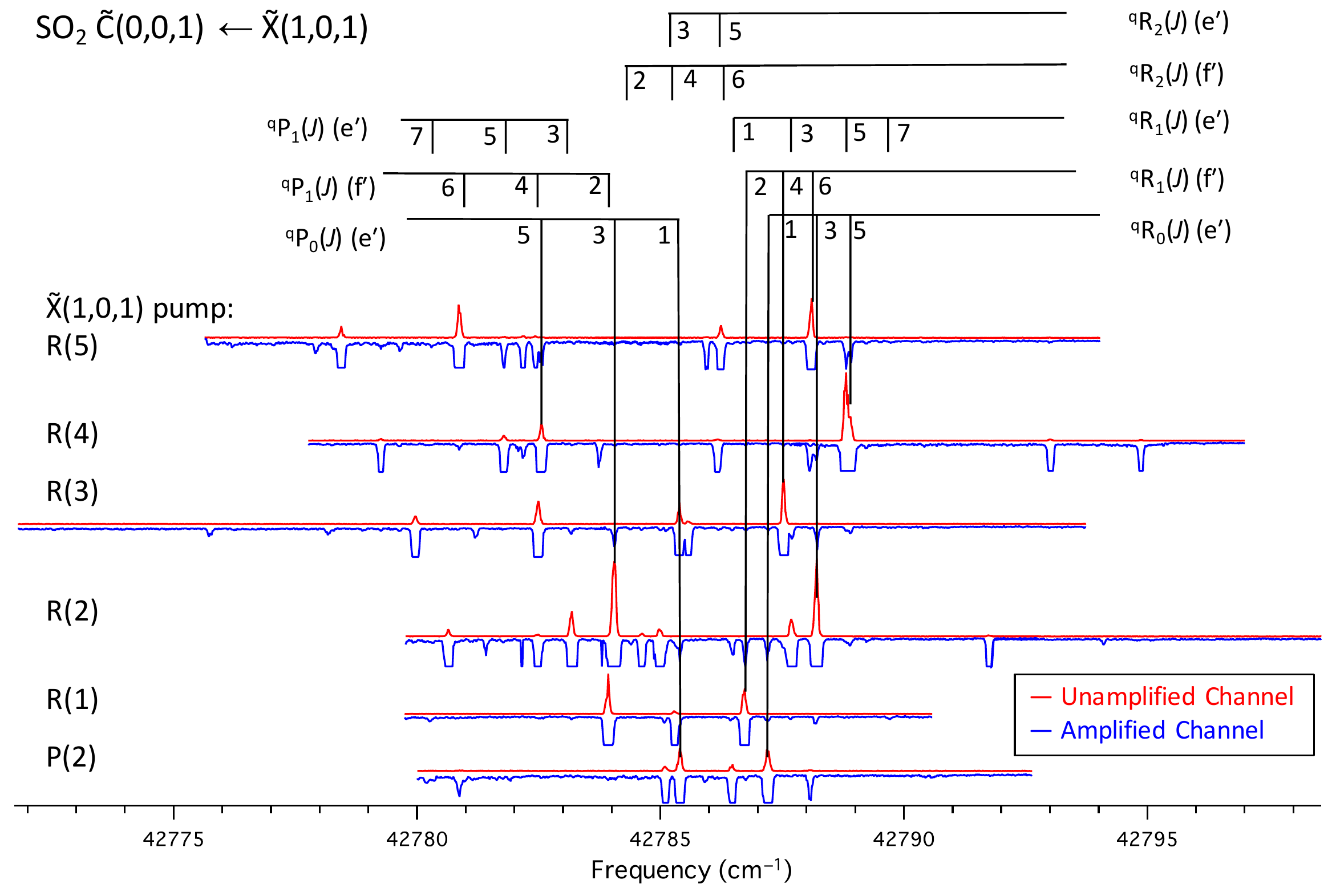}  
\caption{Spectra of the $\mathrm{\tilde{C}}(0,0,1)\leftarrow\mathrm{\tilde{X}}(1,0,1)$ transition, observed by IR-UV double resonance. The energy of the $\mathrm{\tilde{X}}(1,0,1)$ origin (2499.87 cm$^{-1}$) has been added to the frequency axis. The unamplified signal channel is shown as upward peaks (red) and the amplified signal channel is shown as downward peaks (blue). The main $^{\mathrm{q}}$P and $^{\mathrm{q}}$R branches are labeled, and the e$'$ and f$'$ labels refer to the e/f symmetry of the upper state. For simplicity of presentation, the weaker $^{\mathrm{q}}$Q branches and $\Delta K_a=2$ transitions are not labeled. The IR pump transition used in each trace is shown on the left-hand side. There were typically several $K_a$ levels populated at each IR pump transition frequency (see Table S.II of the supplementary material).\cite{SO2_IRUV_1_Supplement}
}
\label{C001FigureLabeled}
\end{figure*}

The IR-UV spectrum to the $v_3'=1$ level is shown in Figure \ref{C001FigureLabeled}. The rotational structure of the $\nu_3$ fundamental band of the $\mathrm{\tilde{C}}$ state does not appear to be significantly perturbed. The inertial defect of 0.862 amu$\cdot$\AA$^2$ is similar to that of the zero-point level (0.361 amu$\cdot$\AA$^2$) and the average error obtained in fitting the rotational structure (0.017 cm$^{-1}$) is smaller than the laser calibration uncertainty, so we conclude that for the $J$ and $K_a$ values observed in our jet cooled spectra, there are no significant Coriolis interactions. The nominally allowed interaction via Eq.\ (\ref{Coriolis23}) is with the (0,1,0) level, which lies 165 cm$^{-1}$ higher in energy, so we do not expect any significant interactions at low rotational quanta via the $t_1^{(2)}\approx0.3$ cm$^{-1}$ $c$-axis Coriolis matrix element. 

We note that the estimate of the $\nu_3'$ fundamental made by Hoy and Brand\cite{BrandSO2MolPhys} (212 cm$^{-1}$) is in remarkably good agreement with our observation (212.576 cm$^{-1}$). The quantitative accuracy of their prediction appears to be the result of a fortuitous cancellation between the cross-anharmonicity between $\nu_2$ and $\nu_3$ (neglected in Ref.\ \onlinecite{BrandSO2MolPhys}) and the error in the inferred location (via perturbations in the $\mathrm{\tilde{C}}$(0,0,2) level) of the $\mathrm{\tilde{C}}$(0,1,1) level, which was too high by $\sim$10 cm$^{-1}$. The incorrect value of 228 cm$^{-1}$ for the $\nu_3'$ fundamental reported by Ivanco\cite{IvancoThesis} appears to have been obtained from a misassigned feature in his hot-band fluorescence spectrum. 

\subsection{The $\mathrm{\tilde{C}}$(0,1,1) and $\mathrm{\tilde{C}}$(0,0,2) levels}\label{section011}
Hallin noted a large inertial defect and a large average error in his fit to the effective rotational constants of $\mathrm{\tilde{C}}(0,0,2)$ (see Table \ref{effrotconstants}) and surmised that the perturbed rotational structure was the result of Coriolis interactions with $\mathrm{\tilde{C}}(0,1,1)$, which had not been directly observed at that time. He performed a fit to the (0,0,2) rotational structure to obtain the term energy of (0,1,1) and its $t_1^{(2)}$ matrix element for $c$-axis Coriolis interaction with (0,0,2).\cite{HallinThesis} (See Table \ref{fit011002}.) We have reanalyzed the interaction using Hallin's data and our direct observations on (0,1,1). Because Hallin's spectrum was recorded at dry ice temperature ($T=195$ K), while our spectrum was recorded in a supersonic jet expansion ($T_{\mathrm{rot}}\approx10$ K), the data set for (0,0,2) extends significantly higher in $J$ and $K_a$ than does our data set for (0,1,1). As a result, our attempts to fit to both the centrifugal distortion constants and the Coriolis matrix element simultaneously led to large correlations between the parameters. Therefore, we constrained the $\Delta_J$ and $\delta_J$ centrifugal distortion constants of (0,0,2) and all quartic centrifugal distortion constants of (0,1,1) to reasonable values (those of the $\mathrm{\tilde{C}}(0,0,0)$ level). Even with this constraint in place, significant correlation ($>$\,0.99) remained between the $C$ constants and the $t_1^{(2)}$ matrix element. We therefore used the inertial defects to impose a constraint on the $C$ rotational constants. 

The inertial defect, $\Delta=I_c-I_a-I_b$, should be zero for a planar rigid rotor. However, in a vibrating, rotating planar molecule, it is different from zero as a result of Coriolis, vibrational, centrifugal distortion, and electronic contributions:
\begin{equation}\label{InertialDefectTerms}
\Delta=\Delta_{\mathrm{Cor}}+\Delta_{\mathrm{vib}}+\Delta_{\mathrm{CD}}+\Delta_{\mathrm{elec}}.
\end{equation}
Harmonic expressions for the last three terms of Eq.\ (\ref{InertialDefectTerms}) were derived by Oka and Morino for triatomic C$_{2v}$ molecules.\cite{OkaTriatomicInertialDefect} 
In the $\mathrm{\tilde{C}}$ state of SO$_2$, we expect the electronic contribution to be negligible, and we expect the contribution of $\Delta_{\mathrm{CD}}$ to be approximately an order of magnitude smaller than the contribution of $\Delta_{\mathrm{vib}}$. 
Hallin assumed that $\Delta_{\mathrm{Cor}}$ was the primary contribution to Eq.\ (\ref{InertialDefectTerms}), so he constrained the $C$ rotational constants with the requirement $\Delta=0$ in order to reduce correlation in his fit between the Coriolis interaction matrix elements and the rotational constants.\cite{HallinThesis} However, the expected values of $\Delta_{\mathrm{vib}}$ for the (0,0,2) and (0,1,1) levels are 0.58 and 0.69 amu$\cdot$\AA$^2$, respectively, so the $\Delta_{\mathrm{vib}}$ contributions are not negligible relative to the contributions of $\Delta_{\mathrm{Cor}}$ (approximately 4.05 and $-2.56$ amu$\cdot$\AA$^2$, respectively). Therefore, we obtain smaller fitting error and better determination of fit parameters when we instead constrain the $C$ rotational constants with the requirement $\Delta=\Delta_{\mathrm{vib}}$. That is, we constrain the fit so that our deperturbed rotational constants retain the expected contribution of vibration to the inertial defect, but we remove the contribution from Coriolis interactions and centrifugal distortion, which are explicitly included in the fit model. We obtain the value of $\Delta_{\mathrm{vib}}$ from our force field fit, described in Part II of this series.\cite{SO2_IRUV_2}


We ignore effects from the interaction of (0,1,1) with (0,2,0) because the energy denominator is large ($\sim$170 cm$^{-1}$) and is not expected to make a significant contribution to the low-lying rotational energy levels that we observe. The resulting fit, given in Table \ref{fit011002}, to a combined data set of 542 lines has an average error of 0.030 cm$^{-1}$ and no correlations greater than 0.75. The observed origin of $(0,1,1)$ is approximately 10 cm$^{-1}$ lower than the fit value obtained by Hallin, and our observed (0,1,1) $A$ rotational constant is significantly larger. The $A$ constants we obtain for (0,1,1) and (0,0,2) are in better agreement with physical expectations, because we expect the $A$ constant to increase when one quantum of stretching excitation is exchanged for one quantum of bending excitation. 
\begin{table}
\caption{Fit parameters for the interacting levels (0,1,1) and (0,0,2) of the $\mathrm{\tilde{C}}$ state obtained from the current work are compared to the fit parameters of Ref.\ \onlinecite{HallinThesis}. All energies are in cm$^{-1}$ units and the inertial defects ($\Delta$) are listed in amu$\cdot$\AA$^2$ units. Values in parentheses are the $2\sigma$ statistical fit uncertainty of the final significant digit. \label{fit011002}} 
\centering
\resizebox{\linewidth}{!}{   
\begin{tabular}{r@{\extracolsep{8pt}}cc@{\extracolsep{6pt}}cc} \toprule
&\multicolumn{2}{c}{This work}&\multicolumn{2}{c}{Ref.\ \onlinecite{HallinThesis}} \\
\cline{2-3}  \cline{4-5}
&(0,0,2)&(0,1,1)&(0,0,2)&(0,1,1) \\
\colrule 
$T_0$       & 43134.679(8)                                  & 43155.646(10)                       &43134.674(2)    &43166.747(46)   \\
$T_{\mathrm{vib}}$ & 561.229(8)                      & 582.196(10)                            &561.224(2)        & 593.297(46)        \\
$A$           & 1.14432(37)                                     & 1.1695(10)                              &1.144971(77)   &1.12941(62)       \\
$B$           & 0.342894(51)                                   & 0.33821(24)                            &0.343338(19)   &0.34651(13)      \\ 
$C$           & 0.2614585\footnote{Constrained. See text.}& 0.259548\footnotemark[1]&0.264133\footnotemark[1] &0.26516\footnotemark[1]\\
$\Delta$   & 0.581                                               & 0.692                                     & 0                              & 0\\
$\Delta_J\times10^{7}$&4.98\footnotemark[1]&4.98\footnotemark[1]             &4.98\footnotemark[1]  &4.98\footnotemark[1]    \\
$\Delta_{JK}\times10^{7}$&$-162(14)$           &129.2\footnotemark[1]           &81.6(27)                       &129.2\footnotemark[1] \\
$\Delta_{K}\times10^{7}$&273(28)                   &73.8\footnotemark[1]              &192.2(51)                    &73.8\footnotemark[1]   \\
$\delta_{J}\times10^{7}$&1.6\footnotemark[1]&1.6\footnotemark[1]                &1.6\footnotemark[1]  &1.6\footnotemark[1]  \\
$\delta_{K}\times10^{7}$&183(20)       &84\footnotemark[1]                              &90.1(15)                      &84\footnotemark[1]        \\
$t_1^{(2)}$&\multicolumn{2}{c}{0.2978(7)}&\multicolumn{2}{c}{0.40183(13)}  \\
Ave. Error &\multicolumn{2}{c}{0.030} &\multicolumn{2}{c}{0.010}\\
\botrule                         
\end{tabular}
}
\end{table}

\begin{figure}
\centering
\includegraphics[width=\linewidth]{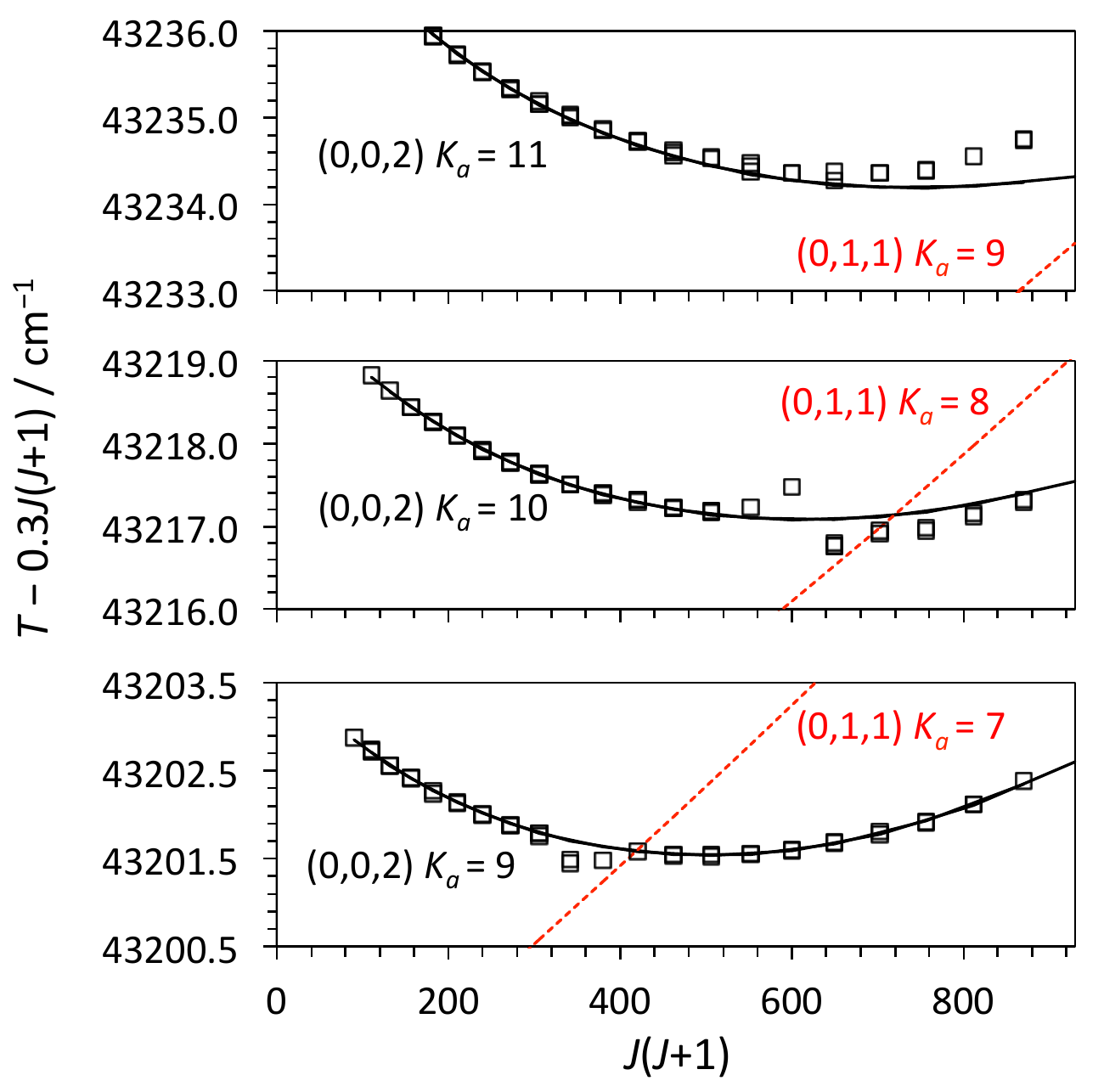}  
\caption{
The term values of the $K_a=9$--11 stacks of the $\mathrm{\tilde{C}}$(0,0,2) state, obtained from Ref.\ \onlinecite{HallinThesis}, are reduced by (0.3 cm$^{-1})\times J(J+1)$ and plotted as open squares vs.\ $J(J+1)$. Our fit to the interacting levels (0,0,2) and (0,1,1) is shown as solid black and dashed red curves, respectively. The upward curvature of each (0,0,2) $K$-stack is due to Coriolis interaction with (0,1,1). An apparent rotationally-dependent local perturbation occurs in the measured (0,0,2) levels. However, the perturbation \emph{cannot} arise from the nearby $\Delta K_a=2$ crossings, because the levels in question have opposite parity. Since there are no other $\mathrm{\tilde{C}}$-state vibrational levels in this region, the interaction most likely comes from a dark perturbing level of a different electronic state.
}
\label{Perturbations002}
\end{figure}

Upon inspection of our fit, we noticed some anomalies in the $K_a=9$--11 stacks of (0,0,2), which resemble a local perturbation. A reduced term value plot in Fig.\ \ref{Perturbations002} highlights the perturbation. The perturbation follows a regular pattern and appears to have a rotational dependence. It appears to affect $J=17$--18 of $K_a=9$ with a magnitude of $\sim$0.2 cm$^{-1}$, $J=23$--27 of $K_a=10$ with a magnitude of $\sim$0.5 cm$^{-1}$, and $J\geq27$ of $K_a=11$ with a magnitude of at least 0.5 cm$^{-1}$. (The data set does not extend past $J=30$.) Coincidentally, the perturbations occur close in energy to predicted levels of (0,1,1) with the same value of $J$, but differing by two units of $K_a$. However, a $\Delta K_a=2$ interaction between (0,0,2) and (0,1,1) is rigorously forbidden because the levels in question have opposite parity. Since there are no other nearby $\mathrm{\tilde{C}}$-state vibrational levels, we conclude that the local perturbation---if it is not an artifact in the data of Ref.\ \onlinecite{HallinThesis}---could only arise due to interaction with a dark level belonging to a different electronic state. Quantum beats have been observed (in the current work) in $\mathrm{\tilde{C}}(0,2,1)$, and have also been reported in a number of higher lying vibrational levels.\cite{Ivanco_SO2_QuantumBeats} An apparent local perturbation in the $J_{K_aK_c}=7_{16}$ rotational level of the higher-lying b$_2$ vibrational level at 45328 cm$^{-1}$, with a magnitude of $\sim$0.4 cm$^{-1}$ has also been observed.\cite{MOMplex} However, we are unaware of local perturbations to other $\mathrm{\tilde{C}}$-state levels that follow similar $J$ and $K_a$ dependence. Therefore, without additional information, we cannot with certainty invoke a specific interaction with a different electronic state. Unfortunately, the original spectrum of $\mathrm{\tilde{C}}(0,0,2)$ photographed in Ref.\ \onlinecite{HallinThesis} is no longer in existence, so we are unable to re-check the assignments. We note that in the fit from Ref.\ \onlinecite{HallinThesis}, which places the (0,1,1) origin too high by 11 cm$^{-1}$, the perturbations occur at weakly avoided $\Delta K_a=3$ crossings with (0,1,1). The shifts at the apparent avoided crossings also increase faster than $[J(J+1)-K(K+1)]^{1/2}$, which makes them appear like an interaction with $\Delta K_a>1$ . It therefore seems that the perturbations had unfortunate consequences for the analysis in Ref.\ \onlinecite{HallinThesis}, where the avoided crossings may have been misinterpreted as Coriolis interactions with (0,1,1), leading to an incorrect determination of the (0,1,1) origin. Since we are unable to identify the origin of the perturbation, we have excluded the perturbed rovibrational levels from the fit given in Table \ref{fit011002}.

\subsection{The $\mathrm{\tilde{C}}$(0,0,3), (0,1,2), (0,2,1), and (1,0,0) levels}
\begin{figure}
\centering
\includegraphics[width=\linewidth]{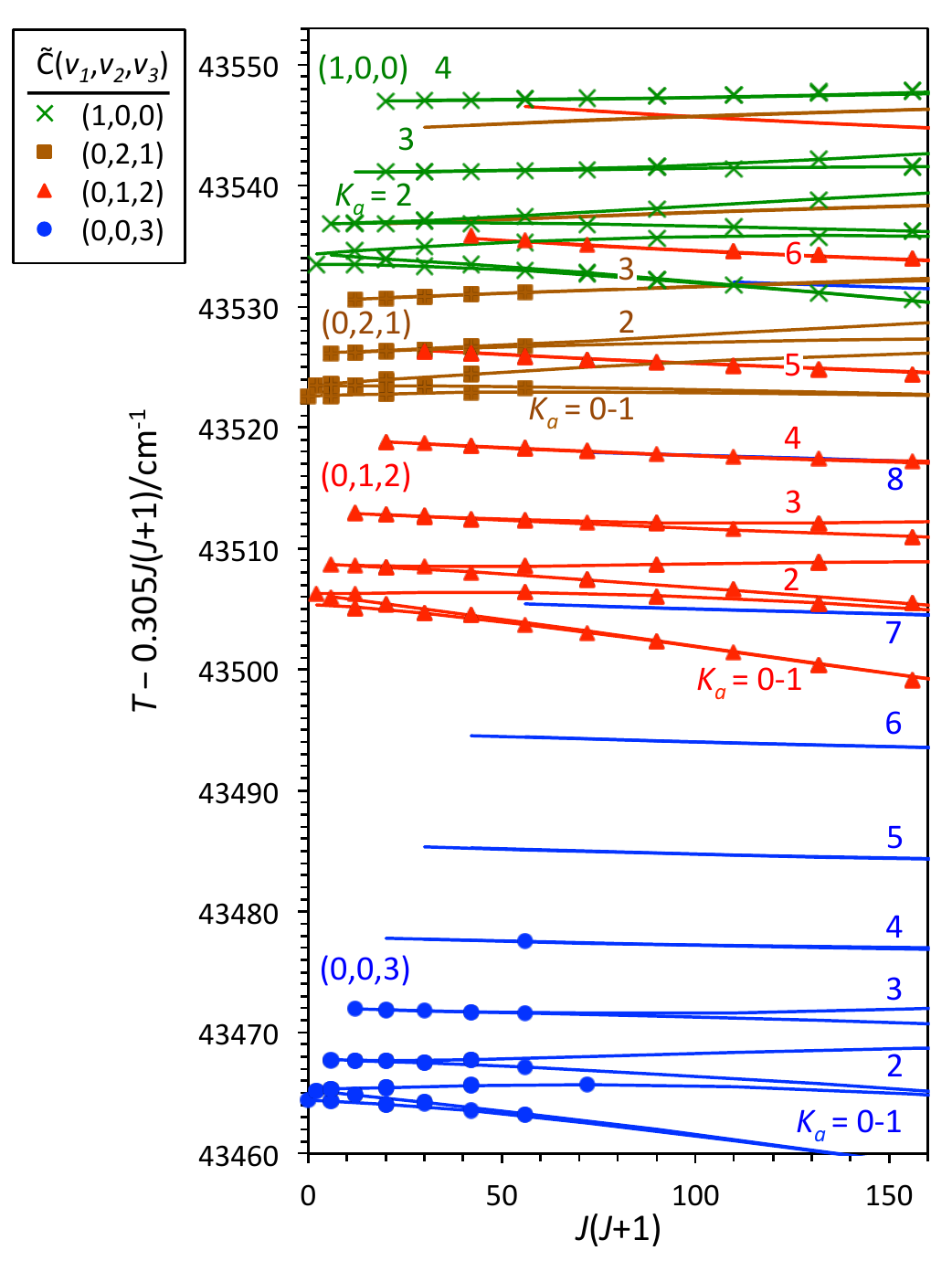} 
\caption{
The observed low-$J$ term energies around 43,500 cm$^{-1}$, reduced by (0.305  cm$^{-1})\times J(J+1)$, are plotted against $J(J+1)$. Curves through the data are from our fit (Table \ref{fittriad}). Experimental data from the (0,0,3) and (0,2,1) levels are obtained in the current work. Data from the (0,1,2) and (1,0,0) levels is taken from Ref.\ \onlinecite{HallinThesis}.
}
\label{triadtermplot}
\end{figure}

The reduced term value plot in Figure \ref{triadtermplot} shows the rotational structure of the $\mathrm{\tilde{C}}$(0,0,3), (0,1,2), (0,2,1), and (1,0,0) levels. The (0,0,3), (0,1,2), (0,2,1), and (0,3,0) levels comprise the next higher set of states that nominally interact via $\zeta_{23}^{(c)}$ interactions. The first three of these states are within $<50$ cm$^{-1}$ of their neighbors, but (0,3,0) is higher in energy by $\sim$170 cm$^{-1}$, and interactions with (0,3,0) are insignificant at low rotational temperatures. Hallin observed the (0,1,2) and (1,0,0) levels and deduced the term value and rotational constants for (0,2,1) by deperturbing the Coriolis interactions.\cite{HallinThesis} Hallin used a model that included not only the $\zeta_{23}^{(c)}$ interaction between (0,1,2) and (0,2,1), but also a higher-order $\zeta_{1/223}^{(c)}$ interaction between (0,2,1) and (1,0,0) and a non-rotationally dependent $t_0^{(4)}$ interaction between (0,1,2) and (1,0,0). 

Our initial attempts to fit simultaneously the origins, rotational constants, and Coriolis interactions to the combined data set were unsuccessful, largely because our jet cooled spectra do not extend as high in $J$ and $K_a$ as Hallin's dry ice temperature spectra. The fit was compromised by interactions occurring at high $J$, $K_a$, preventing the low $J$, $K_a$ data from being fit adequately. However, we believe the interactions are better determined at low $J$, $K_a$, because we have a complete set of observations of the interacting levels. At high $J$, $K_a$, there is significant uncertainty as to the relative contributions from Coriolis interaction and centrifugal distortion to the level structure. Without high $J$, $K_a$ data from the perturbing b$_2$ vibrational levels, it is difficult to disentangle the effects. 

We therefore adopted a two-step scheme for fitting the interactions. In the first step, we fit only the term values with $J\leq 20$ and $K_a \leq 6$. There were large correlations between the rotational constants and the Coriolis parameters, so we again imposed a constraint on the inertial defect, setting the $C$ constants of each level such that $\Delta=\Delta_{\mathrm{vib}}$ obtained from our force field. We then constrained the Coriolis interaction parameters and the rotational constants of (0,0,3) and (0,2,1) to the values obtained from this $J\leq 20$ and $K_a \leq 6$ fit, and we floated the rotational constants of (0,1,2) and (1,0,0) in a fit to the complete data set. The quartic centrifugal distortion constants were constrained to reasonable values. The average error for the (0,1,2) level was unreasonably large, especially at high $J$. We therefore floated the $\Delta_J$ and $\delta_J$ centrifugal distortion parameters for this level and obtained significantly better agreement. 

Our fit results are compared with the fit from Ref.\ \onlinecite{HallinThesis} in Table \ref{fittriad}.  The average error of our combined fit to 871 line positions (0.073 cm$^{-1}$) is significantly larger than the calibration uncertainty ($\sim$0.02 cm$^{-1}$). The average error for the (0,2,1) and (0,0,3) data observed in the current work was reasonable (0.021 and 0.014 cm$^{-1}$, respectively). There is a near degeneracy between the $K_a=5$ stack of (1,0,0) and the predicted location of the interacting $K_a=6$ stack of (0,2,1) that is not reproduced well in our fit, presumably because of uncertainty in the high-$K_a$ level structure of (0,2,1). We therefore omitted the (1,0,0) $K_a=5$, $J\geq 6$ data from our fit and obtained an average error of 0.042 cm$^{-1}$ for the (1,0,0) line positions. 

Most of the fit error comes from the (0,2,1) level, which has an average error of 0.102 cm$^{-1}$. We tried a number of solutions, such as varying the centrifugal distortion constants of (0,1,2) and (0,2,1), and introducing higher order Coriolis interactions. However, none of our attempts to improve the (0,1,2) fit quality led to a well-determined set of physically realistic constants, so we chose not to float any additional parameters in the Hamiltonian. The unexpectedly high value of $\Delta_J$ obtained in our fit may not be physically realistic, but rather partially absorbs other unknown effects not included in our model.

\begin{table*}
\caption{Fit parameters for the interacting levels between 890--960 cm$^{-1}$ of vibrational energy obtained in the current work are compared to the fit parameters of Ref.\ \onlinecite{HallinThesis}. All energies are in cm$^{-1}$ units and the inertial defects ($\Delta$) are listed in amu$\cdot$\AA$^2$ units. Values in parentheses are the $2\sigma$ statistical fit uncertainty of the final significant digit. \label{fittriad}} 
\centering
\resizebox{\textwidth}{!}{ 
\begin{tabular}{rcccc@{\extracolsep{6pt}}ccc} \toprule
&\multicolumn{4}{c}{This work}&\multicolumn{3}{c}{Ref.\ \onlinecite{HallinThesis}} \\
\cline{2-5}  \cline{6-8}
                   &(0,0,3)                                           &(0,1,2)                                      &(0,2,1)                                                &(1,0,0)                                        &(0,1,2)                                        &(0,2,1)                                     &(1,0,0) \\
\colrule 
$T_0$       & 43464.393(11)\footnote{Stated uncertainty is from a fit to the $J\leq20$, $K_a\leq6$ data. The value was constrained in the fit to the full data set.}  
                                                                             & 43505.278(23)                      &43522.566(10)\footnotemark[1] &43533.497(19)                        &43505.429(19)                        &43520.41(23)                        &43533.451(14)\\
$T_{\mathrm{vib}}$ & 890.943(11)\footnotemark[1]& 931.828(23)             & 949.116(10)\footnotemark[1]    & 960.047(19)                            &931.979(19)                             & 946.96(23)                            &960.001(14)\\
$A$           & 1.1432(19)\footnotemark[1]      & 1.16269(33)                           &1.1908(17)\footnotemark[1]          &1.14802(19)                             &1.15721(66)                             &1.2113(70)                            &1.14757(35)\\
$B$           & 0.34049(50)\footnotemark[1]    & 0.33594(27)                           &0.34299(63)\footnotemark[1]       &0.345570(86)                           &0.3308(19)                               & 0.34244(146)                       &0.34514(11)\\ 
$C$           & 0.2595428\footnote{Constrained. See text.} 
                                                                              & 0.2574441\footnotemark[2] &0.2624971\footnotemark[2]        &0.264301\footnotemark[2]      &0.25864\footnotemark[2]      &0.26697\footnotemark[2]      &0.26534\footnotemark[2]\\ 
$\Delta$ (amu$\cdot$\AA$^2$)   & 0.695   & 0.802                                     &0.915                                            &0.316                                        &0                                                &0                                               &0 \\
$\Delta_J\times10^{7}$&4.98\footnote{Constrained to the value for (0,0,2) determined in Ref.\ \onlinecite{HallinThesis}.}
                                                                               &62.0(42)                                  &6.0\footnote{Constrained to the value for $(0,2,0)+\frac{1}{2}[(0,0,2)-(0,0,0)]$ determined in Ref.\ \onlinecite{HallinThesis}.}
                                                                                                                                                                                                &4.98\footnote{Constrained to the fit value determined in Ref.\ \onlinecite{HallinThesis}}
                                                                                                                                                                                                                                                    &5.4\footnote{Constrained to the value for $(0,1,0)+\frac{1}{2}[(0,0,2)-(0,0,0)]$ determined in Ref.\ \onlinecite{HallinThesis}.}  
                                                                                                                                                                                                                                                                                                           &6.0\footnotemark[4]              &4.98\footnote{Constrained to the value for (0,0,0) determined in Ref.\ \onlinecite{HallinThesis}.}   \\
$\Delta_{JK}\times10^{7}$&81.6\footnotemark[3] &95\footnotemark[5]      &109.3\footnotemark[4]                    &66\footnotemark[5]                &95(88)                                        &109.3\footnotemark[4]         &66(7) \\
$\Delta_{K}\times10^{7}$&192.2\footnotemark[3]  &207.5\footnotemark[5]&175.7\footnotemark[4]                   &64\footnotemark[5]                &207.5\footnotemark[6]            &175.7\footnotemark[4]          &64(20)  \\      
$\delta_{J}\times10^{7}$&1.6\footnotemark[3]       &9.7(16)                           &14.2\footnotemark[4]                     &1.6\footnotemark[5]               &7.0\footnotemark[6]                 &14.2\footnotemark[4]            &1.60\footnotemark[7]  \\
$\delta_{K}\times10^{7}$&90.1\footnotemark[3]    &53.0\footnotemark[5]    &97.0\footnotemark[4]                     &84\footnotemark[5]                &53.0\footnotemark[6]               &97.0\footnotemark[4]           & 84.0\footnotemark[7] \\
&\multicolumn{4}{c}{$t_1^{(2)}$[(0,0,3):(0,1,2)$]=0.3250(89)$\footnotemark[1] }                                                                                                       &\multicolumn{3}{c}{$t_1^{(2)}$[(0,1,2):(0,2,1)$]=0.2537(17)$}  \\
&\multicolumn{4}{c}{$t_1^{(2)}$[(0,1,2):(0,2,1)$]=0.3532(44)$\footnotemark[1]}                                                                                                        &\multicolumn{3}{c}{$t_1^{(4)}$[(0,2,1):(1,0,0)$]=-0.0659(16)$}  \\
&\multicolumn{4}{c}{$t_1^{(4)}$[(0,2,1):(1,0,0)$]=-0.0618(32)$\footnotemark[1]}                                                                                                        &\multicolumn{3}{c}{$t_0^{(4)}$[(0,1,2):(1,0,0)$]=1.24(9)$} \\
&\multicolumn{4}{c}{$\textrm{Ave.\ Error}=0.075$}                                                                                                                                                          &\multicolumn{3}{c}{$\textrm{Ave.\ Error}=0.034$} \\
Obs.\ lines:      &50                                        &370                                       &42                                                    &409                                             &370                                             &0                                              &423 \\
\botrule                         
\end{tabular}
}
\end{table*}

Comparing our fit to that from Ref.\ \onlinecite{HallinThesis}, we find Hallin's determination of the (0,2,1) origin to be impressively accurate (within $\sim$2 cm$^{-1}$), considering that no transitions to (0,2,1) had been observed at the time. However, we believe that the Coriolis interaction of (0,1,2) with (0,0,3) makes an important contribution to the rotational structure and cannot be neglected. The likely importance of such an interaction is indicated by the relatively large positive inertial defect in the effective rotational constants of (0,0,3). (See Table \ref{effrotconstants}.) The non-rotationally dependent $t_0^{(4)}$ matrix element reported by Hallin between the (0,1,2) and (1,0,0) levels effectively absorbs some of the effects of Coriolis interaction with (0,0,3). We tried including this perturbation in our fit, but we did not determine it to be statistically different from zero and it did not significantly improve the quality of the fit. The disagreement in the (0,1,2) and (1,0,0) origins between our fit and Hallin's arises from the inclusion of this non-rotationally dependent interaction. One notable difference between the two sets of fit parameters in Table \ref{fittriad} is that our value of the $t_1^{(2)}$ matrix element between (0,1,2) and (0,2,1) is significantly larger than Hallin's. Because the (0,1,2) level is sandwiched between (0,0,3) and (0,2,1), the $c$-axis Coriolis interactions with these two levels have an opposing effect on the effective $C$ rotational constant of (0,1,2). Thus, it is not surprising that the inclusion of (0,0,3) in our fit allows the fitted $t_1^{(2)}$ interaction between (0,1,2) and (0,2,1) to be stronger. Based on harmonic oscillator matrix element scaling arguments (Eq.\ \ref{Coriolis23_1}), we believe our matrix elements are more physically reasonable than Hallin's, because they show the expected increasing trend with increasing vibrational quantum numbers, whereas Hallin's reported matrix element for the (0,1,2):(0,2,1) interaction is \emph{smaller} than the reported matrix element for (0,2,0):(0,1,1). Our determination of the $t_1^{(4)}$ interaction between (0,2,1) and (1,0,0) yields a  value similar to that determined in Ref.\ \onlinecite{HallinThesis}.


\subsection{The $\mathrm{\tilde{C}}$(0,0,4), (0,1,3), (1,0,1), (0,2,2), and (0,3,1) levels}
\begin{figure}
\centering
\includegraphics[width=\linewidth]{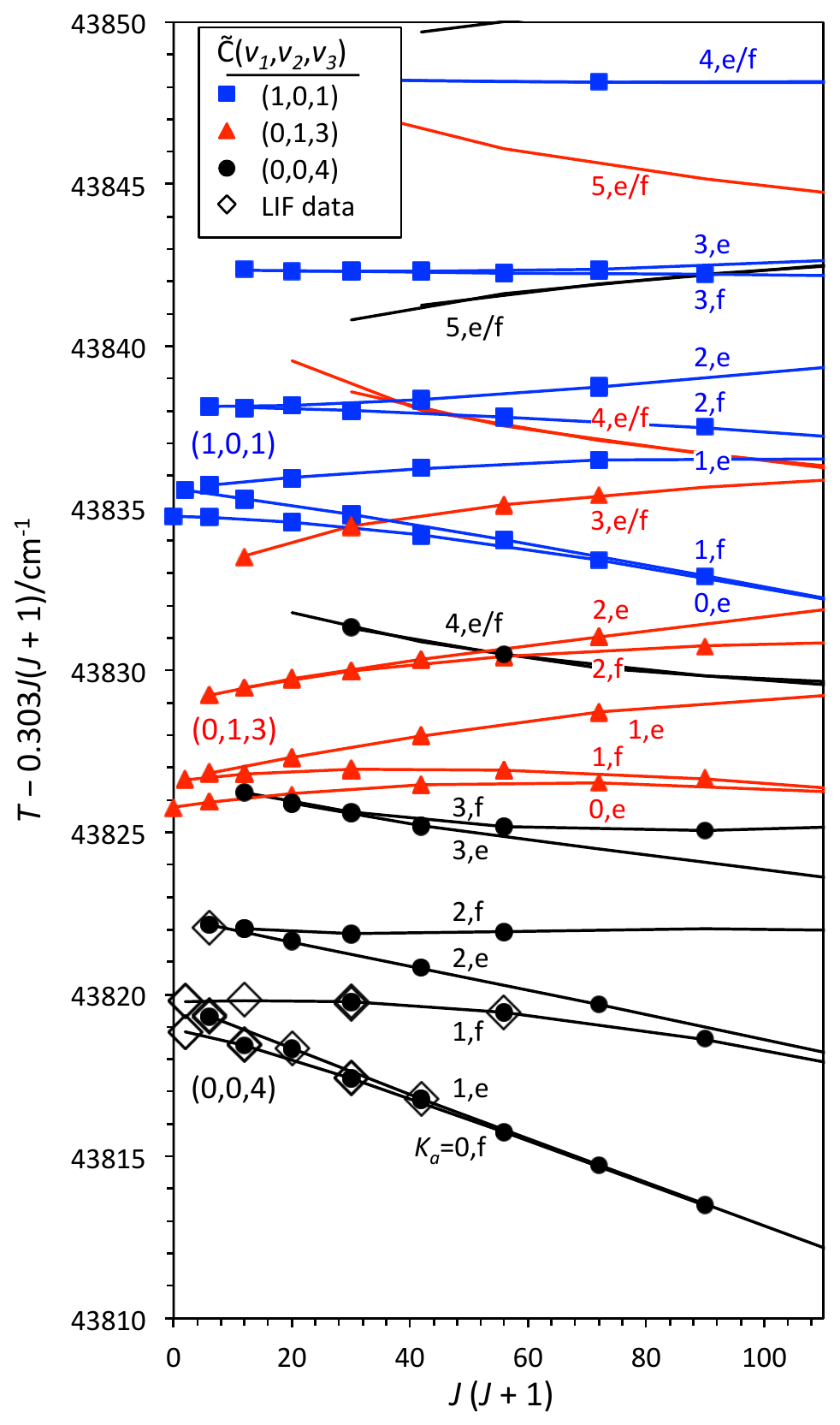}     
\caption{The observed low-$J$ term energies around 43,830 cm$^{-1}$, reduced by (0.303  cm$^{-1})\times J(J+1)$, are plotted against $J(J+1)$. Curves through the data are from the fit shown in Table \ref{fittetrad} and from the (1,0,1) rotational constants given in Table \ref{effrotconstants}. For the (0,0,4) level, experimental data is included from both the one-photon LIF spectrum of Ref.\ \onlinecite{SO2_Yamanouchi} (open diamonds) and from our current double-resonance work (filled circles), where we observe transitions that borrow intensity via $c$-axis Coriolis interactions. 
}
\label{termplot101etc}
\end{figure}

The a$_1$ levels (0,0,4) and (0,2,2) were observed in the one-photon LIF spectrum recorded by Yamanouchi \textit{et al.}\cite{SO2_Yamanouchi} We have observed the nearby b$_2$ levels (0,1,3), (1,0,1), and (0,3,1) via IR-UV double resonance. The reduced term value plots of the lowest three levels in this region are shown in Figure \ref{termplot101etc}.
The origins of the (0,1,3) and (0,0,4) levels are only $\sim$7 cm$^{-1}$ apart and these levels interact strongly via the $c$-axis Coriolis matrix element. As a result, we observe nominally forbidden transitions to (0,0,4) in our IR-UV double resonance spectra. In particular, there is a pathological near-degeneracy between the $K_a>3$ sub-bands of (0,0,4) with the $K_a-1$ sub-bands of (0,1,3). The $K_a=3$ sub-band of (0,0,4) interacts with both $K_a=0$ and $K_a=2$ of (0,1,3). The $K_a=4$ sub-band of (0,0,4) interacts with $K_a=3$ levels of (0,1,3) at low $J$, but is pushed down in energy toward an avoided crossing with $K_a=1$ levels of (0,1,3) at $J\approx10$. Our fit predicts that the $K_a=5$ sub-band of (0,0,4) overtakes the $K_a=4$ sub-band of (0,3,1) so that above $K_a=5$, the energy ordering is reversed and the (nominally) $K_a=5$ sub-band of (0,0,4) is pushed \emph{up} in energy by the Coriolis interaction. However, the levels are strongly admixed, so the nominal labels for (0,0,4) and (0,1,3), $K_a>4$ given in Figure \ref{termplot101etc} are not very good descriptions of the eigenstates. For example, the zero-order $J=5$ levels of the interacting $K_a=4$--5 sub-bands are separated by only 0.26 cm$^{-1}$ and interact with a matrix element of 1.09 cm$^{-1}$, leading to nearly 50:50 admixture of the basis states. Most of our rotational assignments for the (0,0,4) level were not reported in the one-photon LIF study,\cite{SO2_Yamanouchi} presumably because $c$-axis Coriolis perturbations made rotational assignment of the congested LIF spectrum challenging. However, these levels borrow sufficient intensity to be observed and assigned in our IR-UV spectra. We have included data from both the LIF and IR-UV spectra in our fit. 

The (1,0,1) level lies only 10 cm$^{-1}$ higher than (0,1,3), but its rotational structure is not significantly perturbed at low $J$ and $K_a$, as indicated by the small error and moderate inertial defect in the effective rotational constant fit reported in Table \ref{effrotconstants}. We expect any Coriolis interaction between (1,0,1) and (0,0,4) to arise primarily from Fermi admixture of (1,0,1) with (0,1,3) which would then couple (1,0,1) to (0,0,4) via $\zeta_{23}^{c}$, but we do not have sufficient high-$J$ data to observe this higher-order interaction. Therefore, we neglect rotationally dependent interactions with (1,0,1) in our fit. We do, however, find that it is necessary to include all three $\zeta_{23}^{c}$ interactions between (0,0,4), (0,1,3), (0,2,2), and (0,3,1). The highest level in this series of 1:1 $v_2$/$v_3$ exchange would be (0,4,0), which was not observed by Yamanouchi \textit{et al.} due to weak Franck-Condon factors, but it is expected to lie around 44071 cm$^{-1}$, almost 200 cm$^{-1}$ higher than (0,1,3). We therefore do not expect significant Coriolis interactions with (0,4,0) at low $J$. 

The results of our fit to the rotational constants and $c$-axis Coriolis matrix elements for (0,0,4), (0,1,3), (0,2,2), and (0,3,1) are shown in Table \ref{fittetrad}. The $C$ constants and the Coriolis matrix elements were highly correlated (0.999). Therefore, we constrained the $C$ constants in the manner described in Section \ref{section011}, ignoring the $\Delta_{\mathrm{CD}}$ contribution from centrifugal distortion. With this constraint in place, there were no correlations larger than 0.78 in magnitude. 

\begin{table*}
\caption{ Fit parameters for the observed interacting levels with $v_2+v_3=4$. All energies are in cm$^{-1}$ units and the inertial defects ($\Delta$) are listed in amu$\cdot$\AA$^2$ units. Values in parentheses are the $2\sigma$ statistical fit uncertainty of the final significant digit.  \label{fittetrad}} 
\centering
\begin{tabular}{rccccccc} \toprule
                   &(0,0,4)                                           &(0,1,3)                                 &(0,2,2)                                  &(0,3,1)                 \\
\colrule 
$T_0$       & 43818.919(10)                          & 43825.754(14)                  &43873.433(15)                  &43886.691(12)    \\
$T_{\mathrm{vib}}$ & 1245.469(10)                            & 1252.304(14)                     & 1299.983(15)                   &1313.241(12)       \\
$A$           & 1.1389(16)                                & 1.1670(30)                         &1.1861(98)                               &1.2140(25)        \\
$B$           & 0.33983(45)                                   & 0.34035(57)                           &0.3365(11)                          &0.33899(87)          \\ 
$C$           & 0.2585882\footnote{Constrained.}& 0.2601089\footnotemark[1] &0.2580415\footnotemark[1]    &0.2603697\footnotemark[1]          \\ 
$\Delta$ (amu$\cdot$\AA$^2$)   &0.783       &0.834                                       &1.01                                           &1.13                  \\
&\multicolumn{4}{c}{$t_1^{(2)}$[(0,0,4):(0,1,3)$]=0.3463(14)$}   \\
&\multicolumn{4}{c}{$t_1^{(2)}$[(0,1,3):(0,2,2)$]=0.4528(99)$}   \\
&\multicolumn{4}{c}{$t_1^{(2)}$[(0,2,2):(0,3,1)$]=0.4764(42)$}   \\
&\multicolumn{4}{c}{$\textrm{Ave.\ Error}=0.021$} \\
\botrule                         
\end{tabular}
\end{table*}

\subsection{The $\mathrm{\tilde{C}}$(0,0,5) level}
\begin{figure}
\centering
\includegraphics[width=\linewidth]{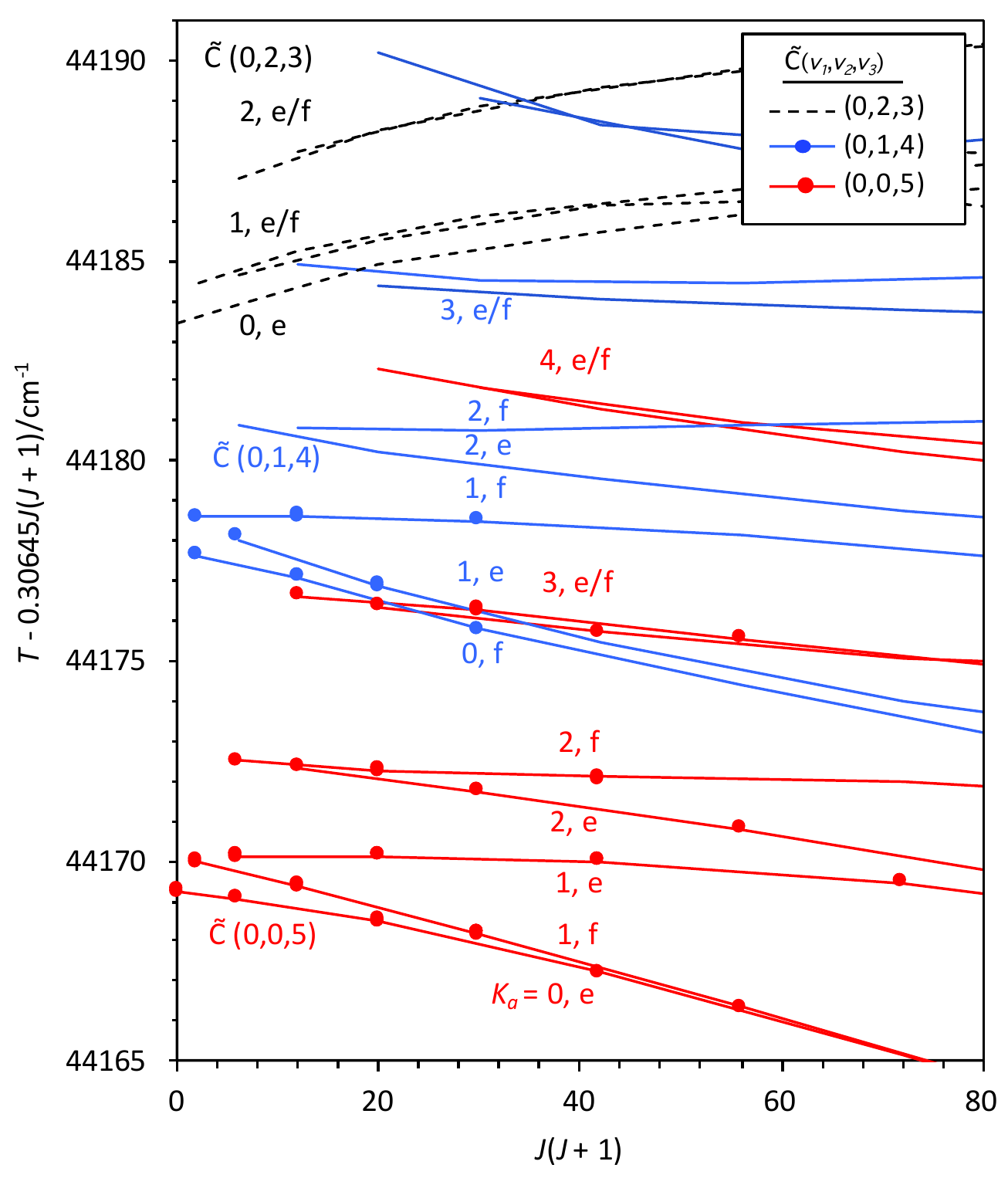}  
\caption{The observed low-$J$ term energies around 44\,180 cm$^{-1}$, reduced by (0.30645  cm$^{-1})\times J(J+1)$, are plotted against $J(J+1)$. Curves through the data are from the fit given in Table \ref{fit005}. Experimental data for the (0,1,4) level is taken from Ref.\ \onlinecite{SO2_Yamanouchi}. The interacting (0,2,3) level has not been directly observed, but its predicted energy from the fit is shown with dotted curves. 
}
\label{termplot005}
\end{figure}

A reduced term value plot of the observed (0,0,5) rotational levels is shown in Figure \ref{termplot005}. The (0,0,5) level has a strong $c$-axis Coriolis interaction with (0,1,4). The (0,1,4) rotational levels reported by Yamanouchi \textit{et al.}\cite{SO2_Yamanouchi} are also shown in the figure. We attempted a fit to the rotational constants and Coriolis matrix element between (0,0,5) and (0,1,4). Although we can fit the data to within $\sim$0.025 cm$^{-1}$, the fit parameters we obtain are unreasonable for several reasons. Both (0,0,5) and (0,1,4) exhibit unreasonably small $C$ constants (0.238 and 0.161 cm$^{-1}$, respectively) and large positive inertial defects (6.40 and 41.8 amu$\cdot$\AA$^2$, respectively). Furthermore, the $t_1^{(2)}$ matrix element that we obtain (0.20 cm$^{-1}$) was much smaller than expected. Placing a constraint on the inertial defects leads to divergence of the fit. It is therefore necessary to include (0,2,3) and its $c$-axis Coriolis interaction with (0,1,4). We did not record the spectrum of this level, because it falls just outside of the range that we observed. However, we are able to predict that the (0,2,3) origin lies at 1610 cm$^{-1}$ using our force field (described in Part II of this series) and we predict a reasonable set of rotational constants for (0,2,3) by adding the rotational constants of (0,1,3) and (0,2,0) and subtracting those of (0,1,0). We constrain the origin and rotational constants of (0,2,3) but float the Coriolis matrix element with (0,1,4). We also constrain the $C$ constants of (0,0,5) and (0,1,4) so that $\Delta=\Delta_{\mathrm{vib}}$, as described in Section \ref{section011}. The parameters obtained from our fit are listed in Table \ref{fit005} and the fit is shown in Figure \ref{termplot005}. Despite the relatively strong Coriolis interactions, we do not observe any a$_1$ levels via intensity borrowing in our IR-UV spectra. 
One noticeable feature in the reduced term value plot is a weakly avoided $\Delta K_a=3$ crossing between the $K_a=0$ sub-band of (0,1,4) and the $K_a=3$ sub-band of (0,0,5), which leads to a slight distortion in the rotational structure at $J=5$. Our fit parameters predict severe interactions between the $K_a=4$ sub-band of (0,1,4) and the $K_a=3$ sub-band of (0,2,3), due to a near degeneracy, but this is sensitive to the exact (0,2,3) origin, which has not been measured. 

\begin{table}
\caption{ Fit parameters for the observed interacting levels (0,0,5), (0,1,4), and (0,2,3). The (0,2,3) level has not been observed, but its origin and rotational constants are predicted and the $c$-axis Coriolis matrix element is determined from a fit to the perturbed rotational structure of the observed levels.  All energies are in cm$^{-1}$ units and the inertial defects ($\Delta$) are listed in amu$\cdot$\AA$^2$ units. Values in parentheses are the $2\sigma$ statistical fit uncertainty of the final significant digit.
 \label{fit005}} 
\centering
\begin{tabular}{rccc} 
\toprule
                   &(0,0,5)                                           &(0,1,4)                                 &(0,2,3)       \\                           
                   \colrule 
$T_0$       & 44169.245(13)                          & 44177.730(23)                  &44184.95\footnotemark[1]     \\
$T_{\mathrm{vib}}$ & 1595.795(13)                            & 1604.280(23)                     & 1611.50\footnotemark[1]           \\
$A$           & 1.1399(27)                                & 1.161(26)                         &1.18788\footnotemark[1]         \\
$B$           & 0.33842(84)                                   & 0.3400(25)                           &0.33875\footnotemark[1]         \\ 
$C$           & 0.257394\footnote{Constrained.}& 0.258845\footnotemark[1] &0.26000\footnotemark[1]       \\ 
$\Delta$ (amu$\cdot$\AA$^2$)   &0.891       &1.02                                      &0.882                                            \\
&\multicolumn{3}{c}{$t_1^{(2)}$[(0,0,5):(0,1,4)$]=0.2957(35)$}   \\
&\multicolumn{3}{c}{$t_1^{(2)}$[(0,1,4):(0,2,3)$]=0.5187(75)$}   \\
&\multicolumn{3}{c}{Ave.\ Error $=0.020$} \\
\botrule                         
\end{tabular}
\end{table}

\subsection{Summary of rotational parameters derived in this work}
\begin{figure*}
\centering
\includegraphics[width=\linewidth]{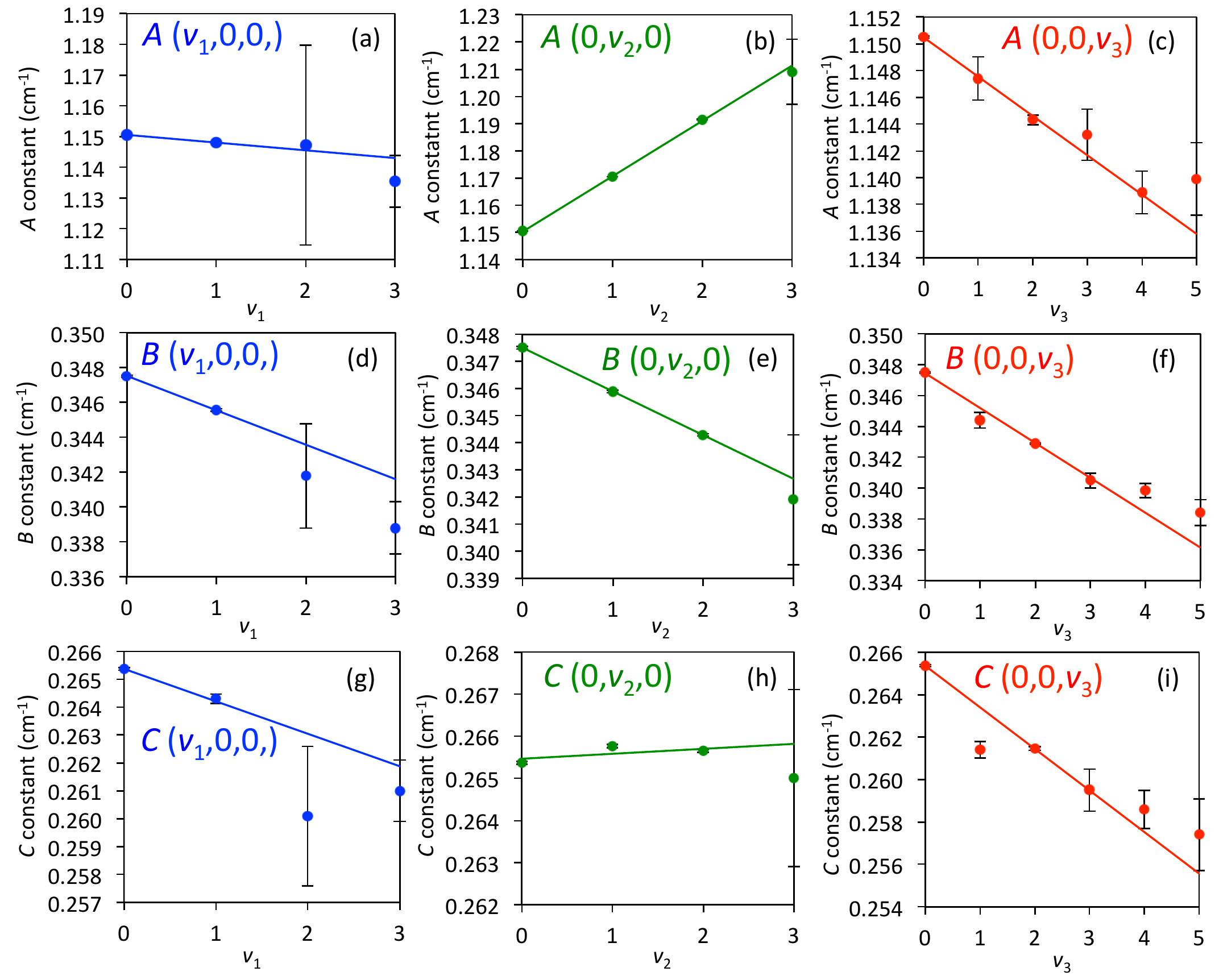}   \\
\caption{The dependence of the deperturbed rotational constants on the vibrational quantum number is plotted for progressions in each normal mode of the $\mathrm{\tilde{C}}$ state of SO$_2$. Error bars represent the statistical 2$\sigma$ standard deviation obtained by fitting the rotational structure, and the weighted linear least-squares fit is shown for each progression. The derived $\alpha$ constants are given in Table \ref{alphatable}.}
\label{alphas}
\end{figure*}

Tables \ref{b2vibterms} and \ref{a1vibterms} provide a summary of the deperturbed vibrational origins and rotational constants obtained from the current work for the low-lying vibrational levels of the $\mathrm{\tilde{C}}$ state of SO$_2$.  In cases where the $C$ constant was constrained via the inertial defects, we estimate the uncertainty in the $C$ constant to be twice the uncertainty in the $B$ constant. 
\begin{table*}
\caption{Term values, deperturbed rotational constants, and inertial defects ($\Delta$) for low-lying b$_2$ levels of the $\mathrm{\tilde{C}}$ state of SO$_2$ below $\sim$1600 cm$^{-1}$ of vibrational excitation. All energies are in cm$^{-1}$ units and inertial defects ($\Delta$) are in amu$\cdot$\AA$^2$ units. Values in parentheses are the $2\sigma$ uncertainty of the final significant digit. \label{b2vibterms}} 
\centering
\vspace{3 pt}
\begin{tabular}{clllllc} \toprule
$(v_1',v_2',v_3')$&  \multicolumn{1}{c}{$T$}&  \multicolumn{1}{c}{$T_{\mathrm{vib}}$}& \multicolumn{1}{c}{$A$} & \multicolumn{1}{c}{$B$} &\multicolumn{1}{c}{$C$}    &\multicolumn{1}{c}{$\Delta$} \\
\colrule 
(0,0,1)  &42786.026(6)    & \phantom{1}212.576(6)   & 1.1474(16)  & 0.3444(5)      & 0.2614(4)      &  0.8644    \\
(0,1,1)  &43155.646(10)  & \phantom{1}582.196(10)& 1.1695(11) & 0.33821(20)   &  0.25955(40)  &  0.6916   \\
(0,0,3)  &43464.393(11)  & \phantom{1}890.943(11) & 1.1432(19) & 0.34049(50)   & 0.2595(10)    &   0.6951   \\
(0,2,1)  &43522.566(10)& \phantom{1}949.116(10)& 1.1908(17)& 0.34299(63)  & 0.2625(13)    &  0.9147    \\
(0,1,3)  &43825.754(14) & 1252.304(14)                      & 1.1670(30)  & 0.34035(57)  & 0.2601(11)    &  0.8343       \\
(1,0,1)  &43834.762(7)   & 1261.311(7)                        & 1.1462(12)  & 0.3420(3)       & 0.2558(2)       &  1.913        \\
(0,3,1)  &43886.691(12)&  1313.241(12)                     & 1.2140(25)   & 0.33899(87)  & 0.2604(17)    & 1.131 \\
(0,0,5)  &44169.245(13)& 1595.795(13)                      & 1.1399(27)  & 0.33842(84)     & 0.2574(17)    &  0.8911         \\
\botrule                         
\end{tabular}
\end{table*}
\begin{table*}
\caption{Term values, deperturbed rotational constants, and inertial defects for low-lying a$_1$ levels of the $\mathrm{\tilde{C}}$ state of SO$_2$ below $\sim$1600 cm$^{-1}$ of vibrational excitation. All energies are in cm$^{-1}$ units and inertial defects ($\Delta$) are in amu$\cdot$\AA$^2$ units. Values in parentheses are the $2\sigma$ uncertainty of the final significant digit. \label{a1vibterms}} 
\vspace{3 pt}
\centering
\begin{tabular}{clllllc} \toprule
$(v_1',v_2',v_3')$&  \multicolumn{1}{c}{$T$}&  \multicolumn{1}{c}{$T_{\mathrm{vib}}$}& \multicolumn{1}{c}{$A$} & \multicolumn{1}{c}{$B$} &\multicolumn{1}{c}{$C$}    &\multicolumn{1}{c}{$\Delta$} \\
\colrule
(0,0,0)\footnote{Ref.\ \onlinecite{HallinThesis}}  &42573.450(4)& \multicolumn{1}{c}{0}       & 1.15050(9)  & 0.34751(5)  & 0.26537(4) & 0.361 \\
(0,1,0)\footnotemark[1]                                             &42950.933(4)& \phantom{1}377.483(4)   & 1.17052(6) & 0.34589(4)  & 0.26576(4) & 0.292  \\
(0,0,2)\footnote{Deperturbed rotational constants obtained from a combined fit to data from Ref.\ \onlinecite{HallinThesis} and data from interacting b$_2$ levels observed in the current work.}
                                                                                      &43134.679(8)& \phantom{1}561.229(8)&1.14432(37)    & 0.342894(51)    & 0.26146(10)    & 0.5811 \\
(0,2,0)\footnotemark[1]                                             &43324.992(4)  & \phantom{1}751.542(4)  & 1.19140(12)&0.34429(4)  & 0.26565(3)  & 0.344 \\
(0,1,2)\footnotemark[2]                                             &43505.278(21)& \phantom{1}931.828(21)&1.16269(31)    &0.33594(27)     &0.25744(54)   &0.8016\\
(1,0,0)\footnotemark[1]                                             &43533.497(19)& \phantom{1}960.047(19)&1.14802(19)&0.345570(86)&0.26430(17)&0.3159           \\
(0,3,0)\footnote{Ref.\ \onlinecite{SO2_Yamanouchi}}&43695.48(3)&1122.03(3)                        &1.209(12)     &0.3419(24)  &0.2650(21)   &0.364      \\
(0,0,4)\footnote{Deperturbed rotational constants obtained from a combined fit to data from Ref.\ \onlinecite{SO2_Yamanouchi} and data from interacting b$_2$ levels observed in the current work.}
                                                                                      &43818.919(10)&1245.469(10)                    &1.1389(16)&0.33983(45)    &0.25859(90)      &0.7831       \\
(0,2,2)\footnotemark[4]                                             &43873.433(15)&1299.983(15)                     &1.1861(98)     &0.3365(11)  &0.2580(22)   &1.012       \\
(0,1,4)\footnotemark[4]                                             &44177.730(23)&1604.280(23)                     &1.161(26)&0.3400(25)  &0.2588(50)   &1.017       \\
\botrule                         
\end{tabular}
\end{table*}
The dependence of the deperturbed rotational constants on the number of vibrational quanta in each normal mode is plotted in Figure \ref{alphas}. Because Ref.\ \onlinecite{SO2_Yamanouchi} assigns only $\sim$10 rotational term values per vibrational state, there is a relatively large uncertainty in the rotational constants for the (2,0,0), (3,0,0), and (0,3,0) states studied in that work. Most of the data points in the third column of Figure \ref{alphas} for the $(0,0,v_3)$ progression were obtained after deperturbation of a manifold of $\zeta_{23}^{(c)}$-type Coriolis interactions, described above. We note that the \emph{effective} (non-deperturbed) $A_{\mathrm{eff}}$, $B_{\mathrm{eff}}$, and $C_{\mathrm{eff}}$ rotational constants shown in Table \ref{effrotconstants} deviate significantly from a linear trend in the $(0,0,v_3)$ progression. The fact that our deperturbed $A$, $B$, and $C$ constants all follow a fairly regular trend with the $(0,0,v_3)$ vibrational progression provides support for the reasonableness of our fitting procedure. The variation in the $A$ rotational constants for the $(0,0,v_3)$ progression may arise from the double-minimum potential energy surface,\cite{Gwinn_pucker1,Gwinn_Pucker2} and the effect is reproduced by our force field.\cite{SO2_IRUV_2} The $C$ constants of (0,1,0) and (0,0,1) are, respectively, higher and lower than the observed trends, partly because of a Coriolis interaction that we have not deperturbed, due to the lack of high-$J$ data from (0,0,1) necessary to make a precise determination of the interaction. (The energy denominator is approximately 165 cm$^{-1}$.)  


The data shown in Figure \ref{alphas} were fit in a weighted linear regression to determine the $\alpha_k^{A,B,C}$ constants, reported in Table \ref{alphatable}. These constants are related to both the molecular structure and to the cubic force field parameters, so accurate determination of the $\alpha_3$ values is important to the characterization of the double-minimum potential. We note also that the deperturbed $C$ constant reported in Ref.\ \onlinecite{HallinThesis} for the (1,0,0) level is nearly identical to the $C$ constant for the (0,0,0) level, even though we expect all of the rotational constants to decrease as symmetric stretching quanta are added. We take this as evidence that the deperturbed constants obtained by Hallin may be distorted due to his choice of fit model, in which the inertial defect was constrained to zero. 


\begin{table}
\caption{The dependence of the rotational constants on quanta of vibrational excitation for the $\mathrm{\tilde{C}}$ state of SO$_2$, 
$\alpha_k^X=-(\partial X/\partial v_k)_0$, where $X=A$, $B$, or $C$ and $k$ is a vibrational mode label. 
Values are listed in cm$^{-1}$ units and numbers in parentheses represent two times the standard error obtained from a weighted fit to the data shown in Figure \ref{alphas}. \label{alphatable}} 
\vspace{3 pt}
\centering
\begin{tabular}{clll} \toprule
$k$&$\alpha^A_k\times10^3$&$\alpha^B_k\times10^3$&$\alpha^C_k\times10^3$           \\
\colrule
  1  & $\phantom{-}2.49$(27)         & 1.98(27)    &$\phantom{-}1.16$(34)   \\
  2  & $-20.4$(5)                           & 1.609(17)& $-0.12(19)$ \\
  3  & $\phantom{-}2.93$(30)     & 2.26(16)  & $\phantom{-}1.95$(27)\\
\botrule                         
\end{tabular}
\end{table}

For the progression in the bending vibration, the $\alpha_2^A$ constant is negative, indicating that the bending vibration experiences a softer potential at the wide bond angle turning point than at the narrow bond angle turning point (because the $A$ constant increases as the geometry approaches linearity). The $\alpha_2^B$ constant is positive because the straightening vibration displaces the oxygen nuclei away from the $b$-axis, but does not move the sulphur nucleus away from the $b$-axis. The $\alpha_2^C$ constant is approximately zero since the straightening vibration moves the oxygen nuclei away from the $c$-axis but moves the sulphur nucleus toward the $c$-axis and the effects approximately cancel. 

Table \ref{CoriolisTable} provides a summary of the $c$-axis Coriolis matrix elements obtained from our fits to the experimental data. For comparison, the predicted matrix elements from a harmonic force field are also listed. Although the experimentally-determined constants scale in a similar fashion as the harmonic prediction, the experimental numbers are all smaller than the harmonic prediction by $\sim$20--50\%, and the discrepancy increases with energy. This discrepancy is not surprising in light of the profoundly \emph{anharmonic} nature of the $\mathrm{\tilde{C}}$ state of SO$_2$. In the $\mathrm{\tilde{C}}$ state, $\nu_2$ is relatively uncoupled, but modes $\nu_1$ and $\nu_3$ are strongly coupled by both Fermi ($K_{1/33}$) and Darling-Dennison ($K_{11/33}$) interactions. Therefore, in the case of $\zeta_{23}^{(c)}$-type interactions among levels with no excitation in $\nu_1$, we expect anharmonic effects to \emph{decrease} the effective $\zeta_{23}^{(c)}$ strength. When there is no excitation in $\nu_1$, Fermi (or Darling-Dennison) resonances cause coupling to levels with one (or two) quanta of $\nu_1$ and $v_3-2$ quanta of $\nu_3$. Because the interacting level has fewer quanta in $\nu_3$, it will have a smaller $\zeta_{23}^{(c)}$ constant and the anharmonic interaction will cause a decrease in the effective $\zeta_{23}^{(c)}$. In Part II of this series,\cite{SO2_IRUV_2} effective Coriolis matrix elements between the anharmonic states are calculated from our force field and compared to the experimental values.


\begin{table}
\caption{Experimentally determined $c$-axis Coriolis matrix elements (in cm$^{-1}$ units) between pairs of interacting levels are compared with the harmonic prediction obtained using the parameters $C=0.26537$ cm$^{-1}$, $\omega_2=392$ cm$^{-1}$, $\omega_3=572$ cm$^{-1}$, and $\zeta_{23}^{(c)}=0.93$. Values in parentheses are the $2\sigma$ uncertainty of the final significant digit.\label{CoriolisTable}} 
\vspace{3 pt}
\centering
\begin{tabular}{clcccc} \toprule
                               &\multicolumn{1}{c}{$t_1^{(2)}$} &$C\zeta_{23}^{(c)}\Omega_{23}[v_2(v_3+1)]^{1/2}$  \\
Interacting levels&\multicolumn{1}{c}{Expt.}    & Harmonic   \\
\colrule
(0,1,1):(0,0,2)&0.2978(7)     &   0.3819    \\
(0,0,3):(0,1,2)&0.3250(89)   &   0.4677   \\
(0,1,2):(0,2,1)&0.3532(44)   &   0.5401    \\
(0,0,4):(0,1,3)&0.3463(14)   &   0.5401   \\
(0,1,3):(0,2,2)&0.4528(99)   &   0.6614    \\
(0,3,1):(0,2,2)&0.4764(42)    &   0.6614    \\
(0,0,5):(0,1,4)&0.2957(35)    &   0.6038    \\
(0,2,3):(0,1,4)&0.5187(75)    &   0.7638    \\
\botrule                         
\end{tabular}
\end{table}



\section{Vibrational level structure}

In the observed vibrational origins given in Tables \ref{b2vibterms} and \ref{a1vibterms} (and pictured in Fig.\ 1 of the third part of this series\cite{SO2_IRUV_3}) it is evident that levels with a single quantum of $\nu_3$ are significantly depressed in frequency. 
However, the degree of odd-even staggering rapidly decreases with increasing $v_3$, indicating a low barrier at the C$_{2\mathrm{v}}$ geometry. The degree of staggering increases as quanta of $v_2$ are added, but decreases when one quantum of $v_1$ is added. The vibrational level structure will be discussed in detail in parts II and III of this series, where we report a new  $\mathrm{\tilde{C}}$-state force field\cite{SO2_IRUV_2} and discuss the mechanism of the vibronic distortions.\cite{SO2_IRUV_3}

\section{Conclusions}
We have observed the low-lying b$_2$ vibrational levels of the $\mathrm{\tilde{C}}$ $^1$B$_2$ state of SO$_2$ via high-resolution IR-UV double resonance spectroscopy. For the low-$J$ rotational levels observed in the supersonic jet expansion, all rotational structure can be fit using an asymmetric top Hamiltonian that incorporates perturbations from $c$-axis $\zeta_{23}$ Coriolis interactions. In cases where high-$J$ data from interacting a$_1$ vibrational levels was available, quartic centrifugal distortion terms are also included. After deperturbing the Coriolis interactions, the vibration-rotation constants are determined. With increasing vibrational energy, the effective Coriolis interaction matrix elements increasingly disagree with the harmonic scaling prediction, largely because of strong Fermi and Darling-Dennison interaction between $\nu_1$ and $\nu_3$. 

The measurements and rotational analyses reported here enable a much more detailed understanding of the vibronically distorted $\mathrm{\tilde{C}}$ state of SO$_2$. Although the staggered $\nu_3$ pattern had been inferred in prior work,\cite{CoonSO2,Brand_SO2_Cstate,BrandSO2MolPhys} and the locations of some of the dark b$_2$ vibrational levels had been estimated from the Coriolis perturbations to bright a$_1$ levels,\cite{HallinThesis,SO2_Yamanouchi} our work has shown that such analyses have sometimes led to inaccurate conclusions about the dark perturbing levels. For example, in the analysis of Ref.\ \onlinecite{HallinThesis}, the energy of (0,1,1) was overestimated  by 11 cm$^{-1}$, and the failure to include interactions with (0,0,3) led to significant errors in the determination of the deperturbed rotational constants of (0,1,2). The observation of b$_2$ vibrational levels allows us to determine much more accurate rotational constants for both the b$_2$ and a$_1$ vibrational levels, because it enables direct deperturbation of the strong Coriolis interactions. In Part II of this series,\cite{SO2_IRUV_2} we describe an internal force field determination of the potential energy surface for the $\mathrm{\tilde{C}}$ state around equilibrium, and in Part III\cite{SO2_IRUV_3} we model the vibronic mechanism for the observed staggered vibrational level structure.

\section{Acknowledgments} \label{Acknowledgments}
The authors thank Anthony Merer and Laurie Butler for valuable discussions. We thank Peter Richter for his assistance in setting up the IR-UV experiments. This material is based upon work supported by the U.S. Department of Energy, Office of Science, Chemical Sciences Geosciences and Biosciences Division of the Basic Energy Sciences Office, under Award Number DE-FG02-87ER13671. 

\bibliographystyle{unsrt}
\bibliography{IRUV_Library}

\end{document}